\documentclass[%
aip,
rsi,
reprint,
amsmath, amssymb, floatfix
]{revtex4-1}
\usepackage{graphicx}% Include figure files
\usepackage{dcolumn}% Align table columns on decimal point
\usepackage{amsmath}
\usepackage[mathlines]{lineno}% Enable numbering of text and display math
\usepackage{subfigure} 
\usepackage{multirow}
\usepackage[utf8]{inputenc}
\usepackage[T1]{fontenc}
\usepackage{mathptmx}
\usepackage{etoolbox}
\usepackage{xcolor}
\usepackage{amsmath}
\usepackage{placeins}

\makeatletter
\def\@email#1#2{%
 \endgroup
 \patchcmd{\titleblock@produce}
  {\frontmatter@RRAPformat}
  {\frontmatter@RRAPformat{\produce@RRAP{*#1\href{mailto:#2}{#2}}}\frontmatter@RRAPformat}
  {}{}
}%
\makeatother
\usepackage{hyperref}

\begin{document}

%\preprint{}

\title{Characterization of a hybrid superconducting RF cavity for use in axion searches}

\author{T. Braine}
    \altaffiliation{thomas.braine@pnnl.gov } 
    \affiliation{Pacific Northwest National Laboratory, Richland, Washington 99354, USA}
\author{G. Carosi}
    \affiliation{Lawrence Livermore National Laboratory, Livermore, California 94550, USA}
\author{E. J. Daw}
    \affiliation{University of Sheffield, Sheffield S10 2TN, UK}
\author{N. Du}
  \altaffiliation{du4@llnl.gov }
  \affiliation{Lawrence Livermore National Laboratory, Livermore, California 94550, USA}
\author{S. Durham}
    \affiliation{Lawrence Livermore National Laboratory, Livermore, California 94550, USA}
\author{C. Mostyn}
    \altaffiliation{cfmostyn1@sheffield.ac.uk}
    \affiliation{University of Sheffield, Sheffield S10 2TN, UK}
\author{M. G. Perry}
    \affiliation{University of Sheffield, Sheffield S10 2TN, UK}
\author{S. Posen}
    \affiliation{Fermi National Accelerator Laboratory, Batavia, Illinois, 60510, USA}
\author{P.J. Smith}
    \affiliation{University of Sheffield, Sheffield S10 2TN, UK}
\author{B. Tennis}
    \affiliation{Fermi National Accelerator Laboratory, Batavia, Illinois, 60510, USA}

\date{\today}
\noaffiliation

\date{\today}

\begin{abstract}
We report on the development of a hybrid superconducting radio-frequency (SRF) cavity for axion haloscopes. The Axion Dark Matter eXperiment (ADMX) Sidecar has been a program of ADMX for testing technologies with the potential for improving high frequency cavity haloscopes in the $4-6\,\mathrm{GHz}$ range. A replica of the Sidecar cavity, an aluminum clamshell microwave resonator loaded with a niobium tuning rod coated with $\mathrm{Nb_3Sn}$, was characterized in a dilution fridge through its critical temperature transitions down to $25\,\mathrm{mK}$ and through magnetic fields up to $6\,\mathrm{T}$. The quality factor of five resonant modes of the cavity system were monitored throughout the process, including the TM$_\text{010}$ mode used in axion searches. Distinct superconducting transitions were observed near the expected critical temperatures of $\mathrm{Nb_3Sn}$, pure niobium, and aluminum. The TM$_\text{010}$ mode exhibited a factors of improvement in quality factor of 1.86 and 2.87 through the two niobium transitions, which implied the rod had a zero-field surface resistance of roughly $79\%$ that of copper at cryogenic temperatures. Magnet tests showed a distinct superconducting transition for aluminum, and gradual degradation of quality factor expected for $\mathrm{Nb_3Sn}$ in a mixed superconducting state. This hybrid resonator design has been integrated in the ADMX insert for performing an axion-like particle search. 
\end{abstract}
\pacs{}
\maketitle

\section{Introduction}
    The nature of dark matter has long been considered one of the largest unsolved mysteries in physics.  One compelling candidate for dark matter is the axion, a hypothetical particle that emerges as a consequence of the Peccei-Quinn solution to the Strong CP problem in QCD \cite{Peccei1977June,PhysRevLett.40.223,PhysRevLett.40.279}. Across a broad range of masses, axions could account for some or all of the dark matter within the Universe~\cite{PhysRevD.83.123531,PhysRevD.91.065014,PhysRevD.92.034507,Fleury_2016,Bonati2016,Petreczky_2016,Borsanyi2016,PhysRevLett.118.071802,Klaer_2017,PhysRevD.96.095001,PhysRevLett.124.161103,10.21468/SciPostPhys.10.2.050,Buschmann2022}.
    \subsection{Axion Cavity Haloscopes}
    %... Insert Intro to conventional axion haloscopes and ADMX...
    Axion haloscopes permit direct laboratory searches for axions from our own galactic halo~\cite{PhysRevLett.51.1415,RevModPhys.82.557,RevModPhys.93.015004}. Axion cavity haloscopes consist of a microwave resonator inside a large static external magnetic field. The external magnetic field stimulates Primakoff conversion of axions to microwave photons. When a TM mode of the resonator is tuned to the same frequency as this microwave photon, axion-to-photon conversion is enhanced. Because of the resonant nature of classical axion haloscopes, axion haloscopes are only sensitive to a narrow mass range of axions. This range can be greatly expanded with the addition of tuning elements to the microwave resonator, such as a tuning rod, which can shift the resonant frequency of the cavity. The rate at which an axion haloscope can probe frequencies for axion dark matter is set by the equation\cite{DawThesis}:

    \begin{align}\label{eqn:axion}
    &\frac{d\nu}{dt}=3.43\,\frac{\text{GHz}}{\text{year}} \left(\frac{V}{100\,\text{L}}\right)^2\left(\frac{B}{8\,\text{T}}\right)^4\left(\frac{C}{0.5}\right)^2 \nonumber \\&\left(\frac{g_\gamma}{0.36}\right)^4
    \left(\frac{\rho_a}{0.45\,\text{GeV/cc}}\right)^2\left(\frac{\nu_a}{1\,\text{GHz}}\right)^2\left(\frac{Q_0}{100,000}\right) \nonumber \\&\left(\frac{\beta^2}{(1+\beta)^3}\right)\left(\frac{3.5}{SNR}\right)^2\left(\frac{0.2\,\text{K}}{T_{sys}}\right)^2, %\numberthis
    \end{align}
    where $V$ is the volume of empty space in the cavity, $B$ is the magnitude of the external magnetic field, $C$ is the cavity form factor (the overlap between the cavity resonant mode and the magnetic field), $g_\gamma$ is the model-dependent numerical constant (most commonly set as $-0.97$ for the KSVZ benchmark model\cite{KSVZ1,KSVZ2} and $0.36$ for the DFSZ model\cite{DFSZ1,DFSZ2}) which determines the axion-photon coupling, $\nu_{a}$ is the potential axion frequency being probed, $\rho_a$ is the expected axion density in the cavity, $Q_0$ is the intrinsic quality factor of the cavity mode (used interchangeably with $Q$ throughout this paper), $\beta$ is the coupling coefficient of the antenna to the cavity mode, $T_{sys}$ is the system noise temperature of the receiver, and $SNR$ is the target signal-to-noise ratio. Here, the equation of the scan rate has been scaled to the experimental parameters typical of an axion haloscope such as ADMX~\cite{PhysRevLett.51.1415}. 

    \par Although the haloscope method has proven to be the most sensitive method for probing QCD axions, the effectiveness of a single, copper cavity haloscope is increasingly limited at frequencies above $\sim1\,\rm GHz$. At higher frequencies, the volume of the resonators is diminished, the quantum limited noise temperature is increased, and, most relevant to this work, the quality factor for copper cavities decreases. Combined, these effects cause the scan rate to scale with frequency as, $\frac{d\nu}{dt} \propto \nu^{-14/3}$, resulting in a significant slow-down in data collection. The diminished quality factor of copper cavities is due to an increase in the anomalous skin depth of copper, lowering its conductivity, such that $ Q_0 \propto \nu^{-2/3}$. For this reason, alternative, high conductivity materials, like superconductors, are being pursued for high frequency haloscope searches.
    \subsection{Superconducting Radio Frequency Cavities}
        Superconducting Radio-Frequency cavities are known for achieving the highest quality factors of any RF resonator in the laboratory. These resonators have become a standard for efficiently accelerating particles in cyclotrons and linear accelerators, routinely achieving $Q\approx10^{10}$, whereas a similar sized copper cavity would be limited to $Q<10^{6}$.\cite{CAS02} Within the dark matter community, superconducting cavities have been used to look for dark photons at an incredibly high sensitivity \cite{RaphaelSRF}. The axion community has done feasibility studies of SRF cavities for several materials, particularly NbTi \cite{BraineMultiMode,QUAX1,QUAX2}, $\mathrm{Nb}_3\mathrm{Sn}$ \cite{PosenNbSN}, and YBCO \cite{CAPPYBCO1,CAPPYBCO2,CAPPYBCO3}.
        \par These cavities must operate in very high magnetic fields in order to detect axions, which is a challenge for superconductors due to the breakdown of the Meissner effect. The Meissner effect refers to the expulsion of magnetic fields from a superconductor's interior during its phase transition\cite{Meissner}. This is a result of the superconductor developing a surface magnetization in response to any applied external field. This effect, however, is limited by the strength of the applied field; above a critical field, $H_c$, the material returns to its normal state.
        \par Superconductors can be classified into Type I and Type II. Type I superconductors exhibit a single critical field, $H_c$, which below it perfectly superconducts, and above which it is completely normal. These superconductors, such as Aluminum, are not practical for use in haloscopes, because none possess $H_c > 1\,\mathrm{T}$\cite{TypeIRohlf}.
        \par Type II superconductors, on the other hand, exhibit two critical fields, $H_{c1}$ and $H_{c2}$, between which a transitional superconducting phase occurs\cite{ABRIKOSOV1957,RJABININ1935}. This mixed state is characterized by quantized vortices where magnetic flux penetrates the superconductor and the material within is in the normal state, whereas the surrounding material still remains superconducting\cite{FEYNMAN1955,Onsager1949,London1950}. These flux vortices, called fluxons, each containing a single magnetic flux quanta, $\Phi_0=h/2e$, increase proportionally in number and density as the external magnetic flux is increased, with their relative size being $\approx\pi\lambda_{L}^2$, where $\lambda_{L}$ is the penetration depth of the material. Above $H_{c2}$, the Type II superconductor returns entirely to the normal state. While $H_{c1}$ typically remains on the order of $\mathcal{O}(10^{-1})\,\mathrm{T}$, $H_{c2}$ can be on the order of $\mathcal{O}(10)\,\mathrm{T}$. In the case of NbTi and $\mathrm{Nb}_3\mathrm{Sn}$, this value is approximately $15\,\mathrm{T}$ and $30\,\mathrm{T}$ respectively\cite{TypeIIBlatt}.
        \par Motion of these fluxons is the primary cause of the RF surface surface resistance in SRF cavities\cite{GRfluxonmodel,PompeoSilva}. To minimize this motion, defects in the grain structure are purposefully introduced to 'pin' fluxons in place\cite{Coffey}. Nonetheless, SRF cavities exhibit a degradation in quality factor with magnetic field up to its $H_{c2}$ \cite{Alimenti2019, BraineMultiMode}. This degradation curve can be stretched by picking materials with very high $H_{c2}$ and $T_c$, such as $\mathrm{Nb}_3\mathrm{Sn}$. Since axion searches typically operate with magnets $B_0< 15 \, \mathrm{T}$ it is possible that a $\mathrm{Nb}_3\mathrm{Sn}$ SRF cavity ($H_{c2}\approx30 \, \mathrm{T}$) would still have a quality factor in excess of a conventional copper version. 
\section{Experiment}
    \subsection{ADMX Sidecar}
    \par Sidecar is a test-bed axion haloscope designed to develop and evaluate new axion search technology. It is a separate high frequency haloscope RF receiver that resides within the ADMX insert. The cavity is mounted atop the ADMX main cavity making use of the dilution fridge cooling power and fringe magnetic field of the main experiment. It is approximately $0.61 \, L$ in detection volume with an approximate resonant frequency tuning range of $4-6\,\text{GHz}$ and at full magnet current ($220 \rm\,A)$ is exposed to an average magnetic field of $3.58 \,\rm T$. First commissioned in 2017, it has demonstrated the novel use piezoelectric motors for cavity tuning\cite{BoutanThesis,Sidecar1A} as well as the first Josephson Traveling Wave Parametric Amplifier used in an axion search\cite{Sidecar1C}.
    \par In its most recent iteration, a $\mathrm{Nb}_3\mathrm{Sn}$ coated tuning rod, known for its high critical field and low surface resistance, was fabricated to improve Sidecar's $Q$. The rod was constructed by Fermilab's Superconducting Quantum Materials and Systems (SQMS) Center\cite{SQMS} using a pure niobium substrate and coated with a $\rm \approx 3\,\mu m$ thick $\mathrm{Nb}_3\mathrm{Sn}$ film using their process outlined in Reference \onlinecite{Posen_2017}. This is considered a thick film ($\approx18\lambda_L\approx10^3\xi$) and is treated in the bulk limit.
    \par This rod was expected, assuming ideal superconducting performance, to improve the $Q$ by 20\% to 90\% depending on its tuning angle, based on simulation of the cavity and rod geometric factors (Equation \ref{eqn:hybridkappaQratio}).
    \par During its commissioning in the ADMX insert, $Q$ values were far lower than expected, and it was suspected that RF radiation was leaking out of the clamshell seam.
    \par Additionally, the NbTi JTWPA within the Sidecar RF receiver prevented observation of the $\mathrm{Nb}_3\mathrm{Sn}$ transition at $T_c\approx18\,\mathrm{K}$ as the amplifier would remain normal and 'opaque' until the NbTi RF cables went superconducting at $T_c\approx 9\,\mathrm{K}$. To further complicate proper characterization, the ADMX cryostat could not be well-controlled during the cool-down and the magnet ramp.
    \par To better characterize the superconducting rod in a controlled environment, measurements were made using a separate replica Sidecar cavity system at the University of Sheffield.
    \par A replica $\mathrm{Nb}_3\mathrm{Sn}$ tuning rod was used in this experiment; its only difference from the rod employed in the ADMX insert was that the pure niobium rod was solid, increasing its mass. The rod used in the ADMX insert was hollow in order to minimize the torque necessary to rotate it via piezoelectric motor as well as minimize thermal mass. Because this replica rod was heavier, it was decided that motion control of the rod would not be feasible, but also was unnecessary for the tests. 
    \par
\begin{figure}[htbp]
    \centering
    \subfigure[Assembled replica Sidecar cavity mounted to QSHS cryostat.]{\label{fig:CavityMounted}\includegraphics[ width=0.7\linewidth]{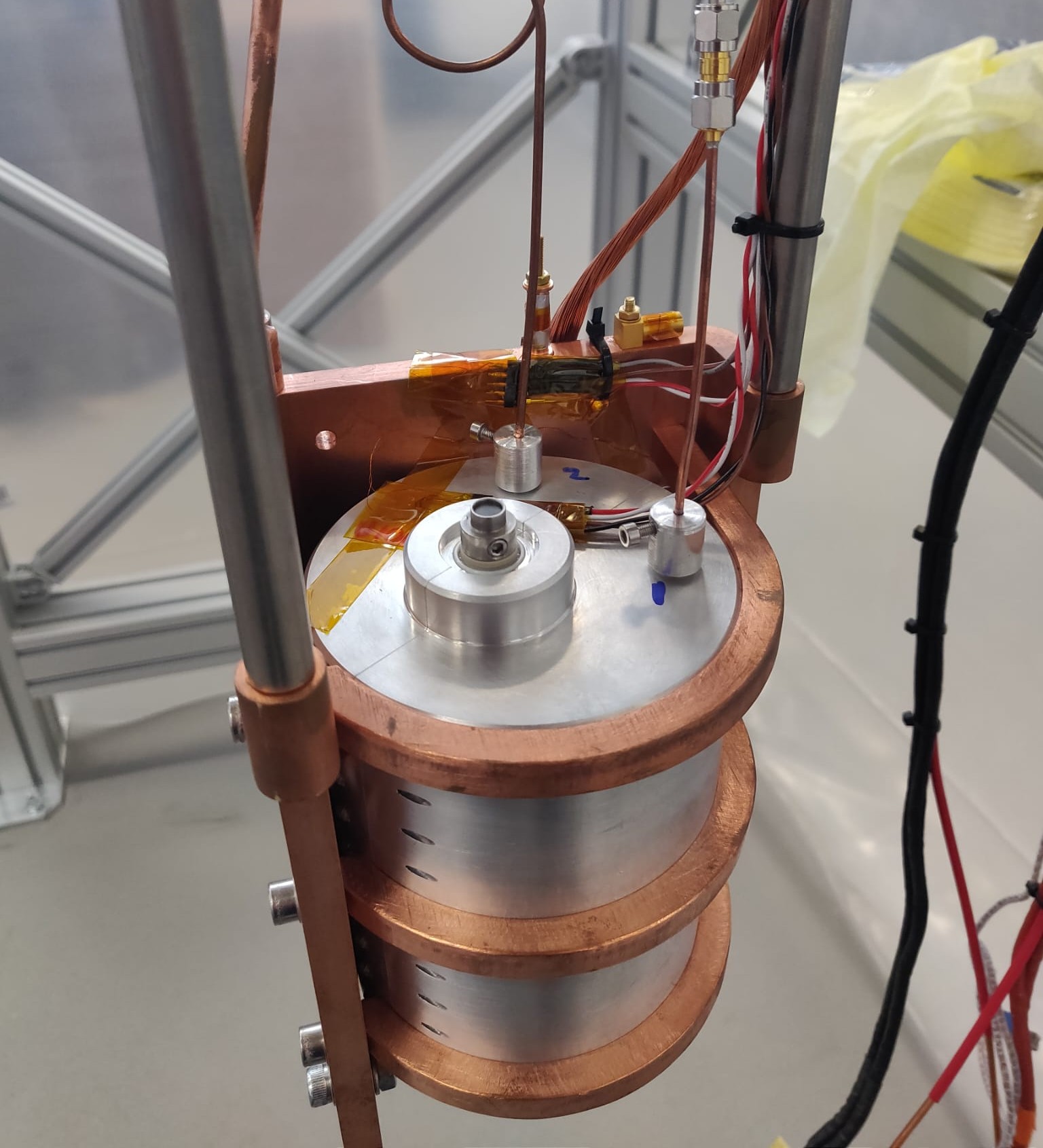}}
    \\
    \subfigure[Bottom view of Sidecar cavity in QSHS cryostat, showing cavity temperature thermometer, mechanical support copper strut, and heat-sink copper braid.]{\label{fig:CavityBottView}\includegraphics[ width=0.7\linewidth]{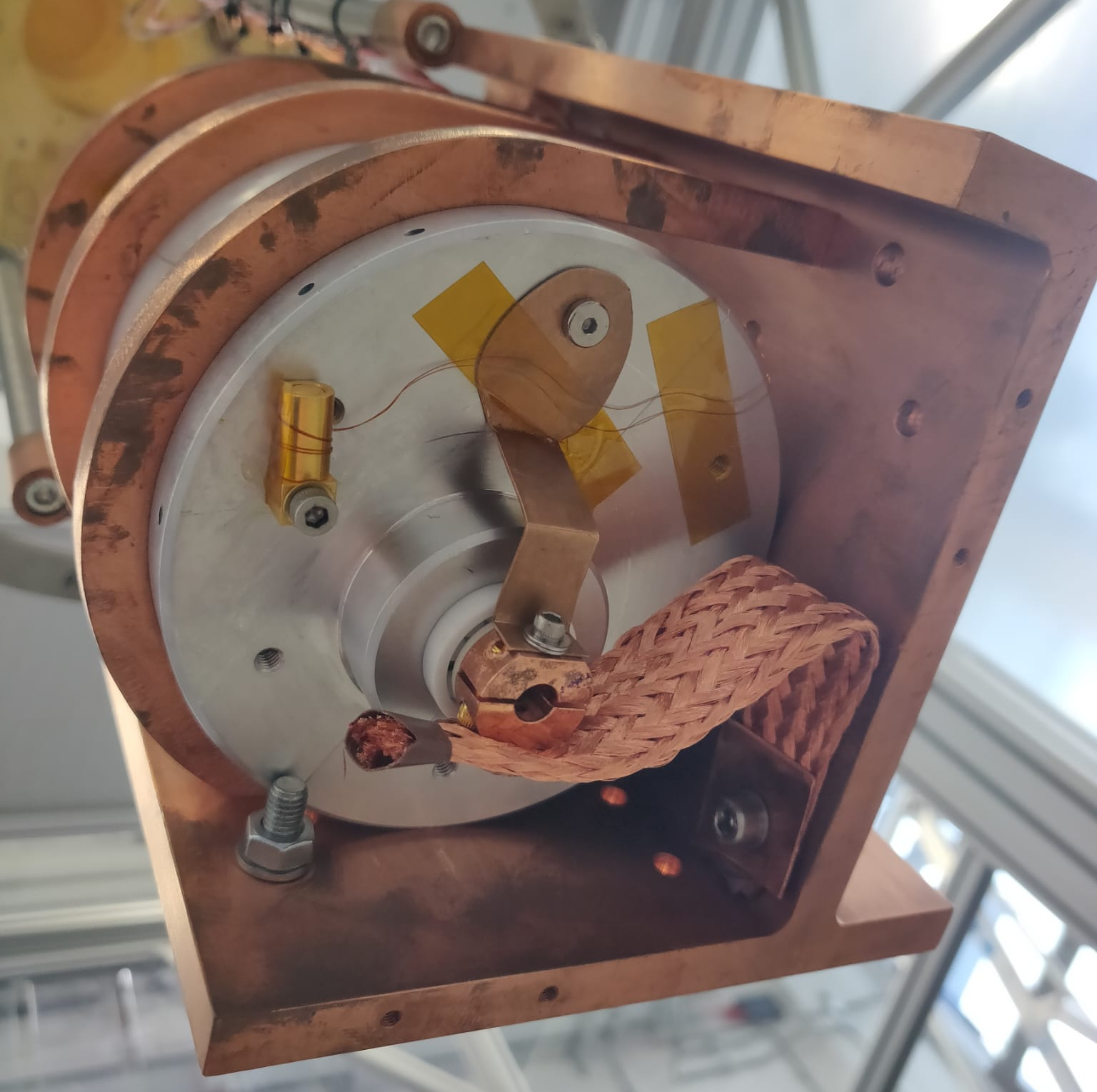}}
    \\
    \subfigure[Aluminum cavity clamshell halve with superconducting tuning rod inserted on its sapphire axles.]{\label{fig:Clamshell}\includegraphics[ width=0.6\linewidth]{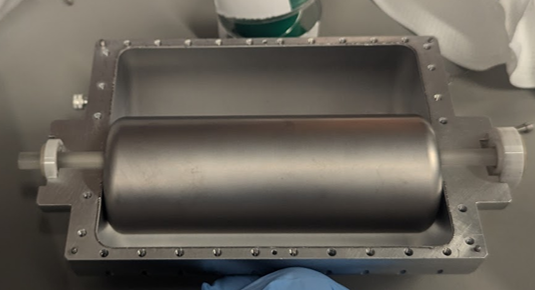}}
    \\
    \subfigure[Assembled superconducting tuning rod with sapphire axle mounted via copper tab inserts.]{\label{fig:RodwTabs}\includegraphics[ width=0.45\linewidth]{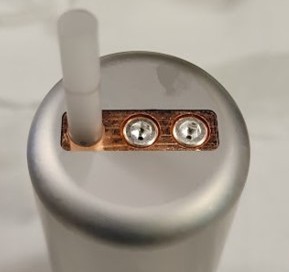}}
    \caption{The Aluminum-$\mathrm{Nb}_3\mathrm{Sn}$ cavity system used in testing.}
    \label{fig:CavityPicsAll}
    \end{figure}

 An aluminum replica Sidecar cavity was used in this study, rather than the copper version installed within the ADMX insert. To minimize any potential radiation losses, an inset seam was machined into each cavity clamshell halve. This seam was then filled with indium to seal each halve of the cavity together. An open clamshell halve with the $\mathrm{Nb}_3\mathrm{Sn}$ coated rod is pictured as well as the cavity assembled and mounted to the cryostat in Figure \ref{fig:CavityPicsAll}.
    \par
\begin{figure*}[htbp]
    \centering
    \includegraphics[width=0.8\linewidth]{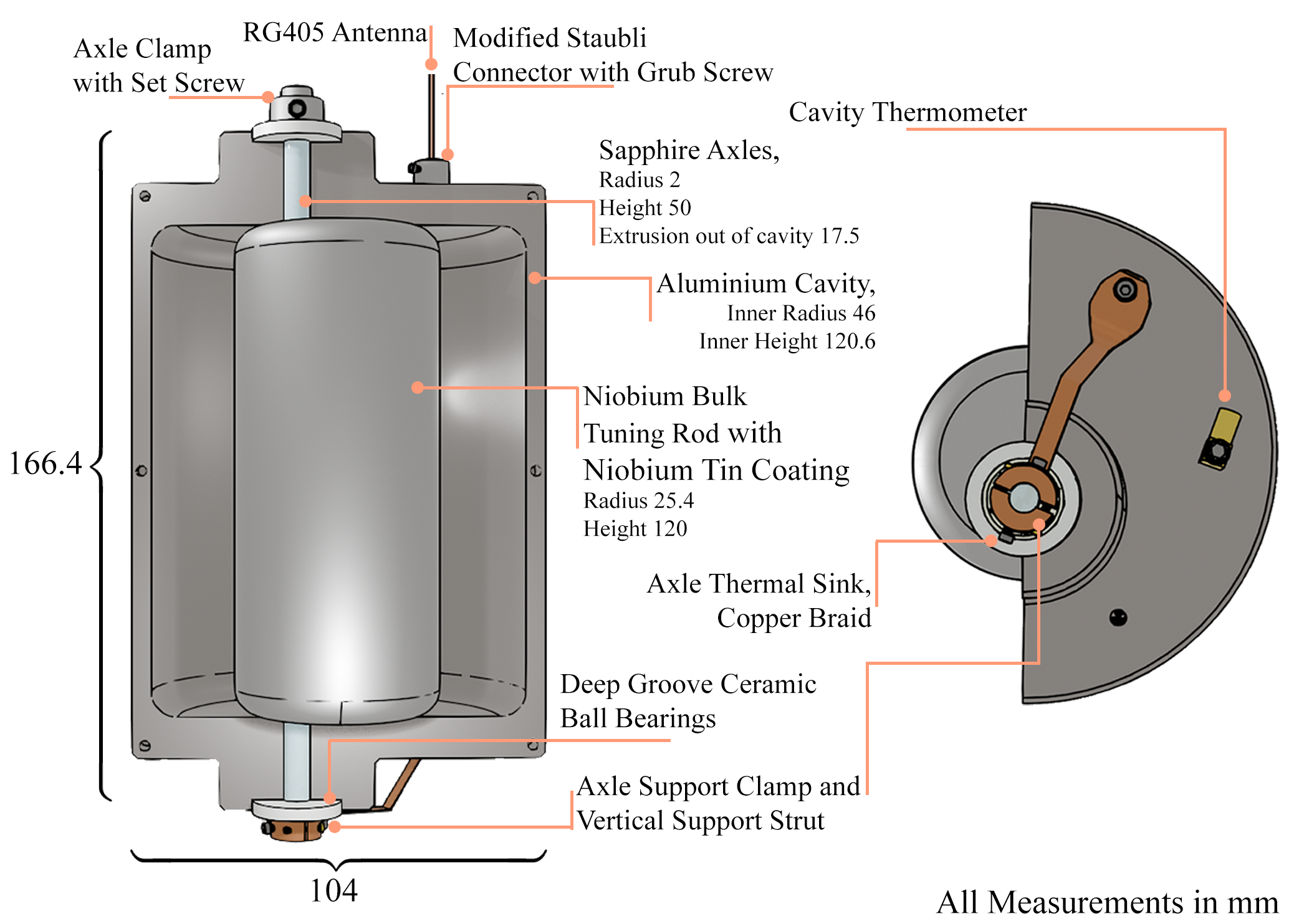}
    \caption{Diagram of the aluminum cavity and $\mathrm{Nb}_3\mathrm{Sn}$ coated tuning rod, indicating how the tuning rod and antennas were held static during the run. A bottom view showing the positions of the cavity thermometer and vertical support strut is shown on the right.}
    \label{fig:CavitySchematic}
    \end{figure*}

 The tuning rod had its sapphire axles attached using copper pieces, referred to in this paper as "axle tabs", that were epoxied on to sapphire with Stycast 2850FT. An inset with threaded holes was machined into each end cap of the rod enabling the copper pieces to be mounted flush with the rod surface using two silver-plated machine screws pictured in Figure \ref{fig:CavityPicsAll}. A full labeled schematic of the cavity and its overall dimensions is shown in Figure \ref{fig:CavitySchematic}.
    
        \subsection{Sheffield Dilution Refrigerator and Magnet}
    \label{sec:QSHSInst}
    
    \par The QSHS apparatus at Sheffield is a dry closed-cycle Proteox MX500 dilution refrigerator (DR) reaching a minimum temperature of $\rm 8.6\,mK$ at the mixing chamber (MXC) plate, and using a superconducting solenoidal magnet with maximum field $6\,\rm T$ (since upgraded to $8\,\rm T$). The DR and magnet were both manufactured by Oxford Instruments (OI, now Quantum Design Oxford) \cite{oxinstproteox}. A full description of the QSHS facility is available in Reference \onlinecite{QSHSInstPaperArx}.
    \par
\begin{figure}[tbp]
    \centering
    \includegraphics[width=1.0\columnwidth]{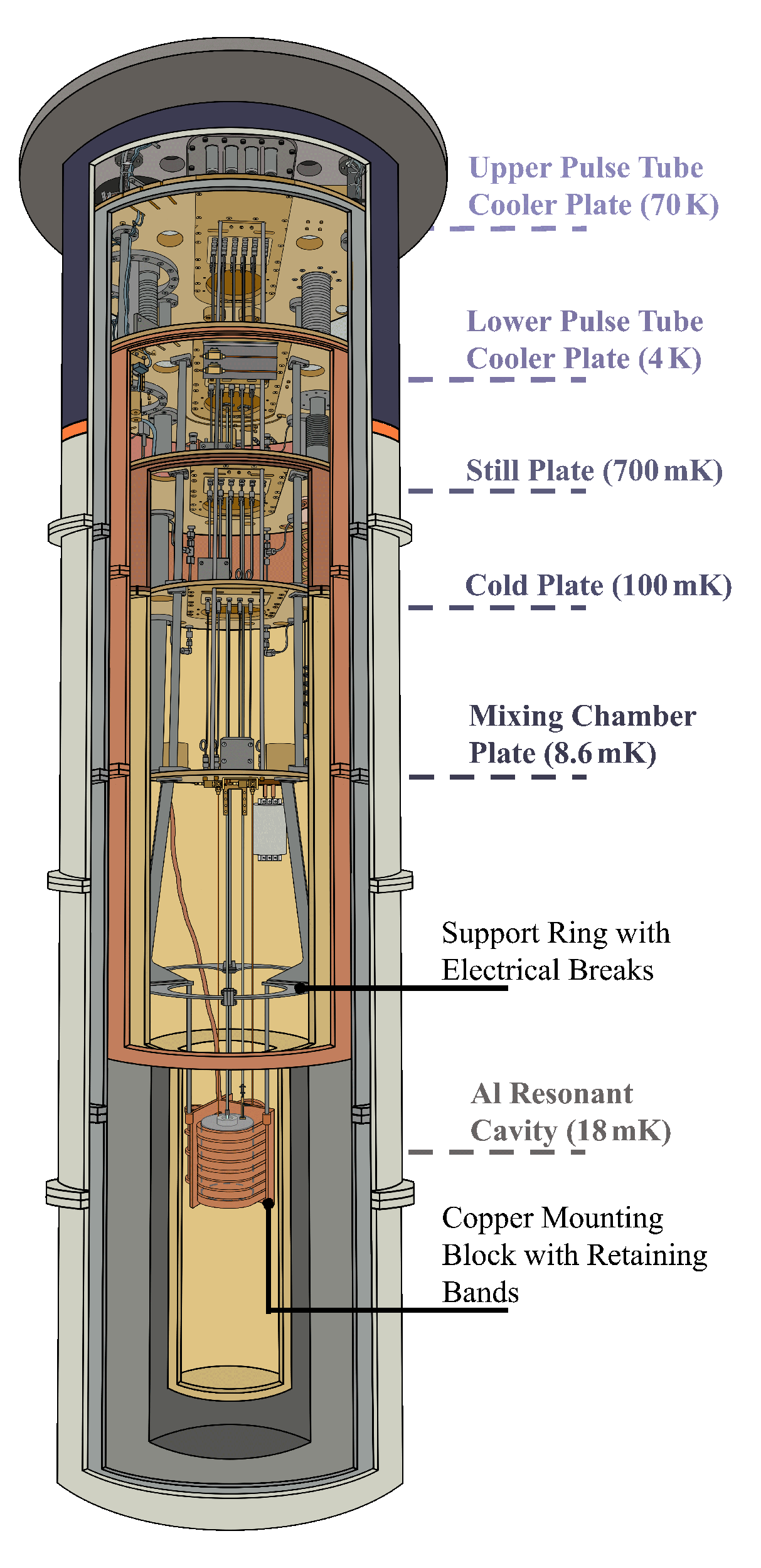}
    \caption{Oxford Instruments \cite{oxinstproteox} Proteox MX500 dilution refrigerator schematic, adapted from \cite{QSHSInstPaperArx}.}
    \label{fig:QSHSDRDiag}
    \end{figure}

 The upper two plates of the DR are cooled using a two stage pulse-tube pre-cooling system and the lower, colder plates using a $^{3}$He/$^{4}$He mixture DR system. During operation, the system is dry and under-vacuum such that no liquid or superfluid helium is present near the internal hardware. An overall schematic of the QSHS DR is adapted from Reference \onlinecite{QSHSInstPaperArx} in Figure \ref{fig:QSHSDRDiag}.

    \par The cavity is mounted  (Figures \ref{fig:CavityMounted} and \ref{fig:QSHSDRDiag}) such that its center is $\rm 677 \pm 5 \,mm$ below the MXC plate. The mounting structure consists of three grade-316 stainless steel (SS) struts extending to a SS circular stiffening ring with G10 insulating spacers used to avoid current loops, further vertical support tubes then extend down into the inner bore of the magnet and connect to a single-piece oxygen-free copper (OFC) mounting block with the cavity secured with curved OFC clamps and kept at fixed vertical position using a resting screw. Thermal sinking to the OFC mount is achieved using an OFC `cold finger': a soft copper rod mounted to both the MXC plate and OFC clamping block. 
    \par Temperature readings are performed using Lakeshore ruthenium oxide (ROX) temperature sensors. Calibrated sensors were supplied on each major plate by OI with further sensors installed by QSHS on the MXC plate, cavity mount, and on the cavity itself, with a specific NIST calibrated ROX sensor used to calibrate these additional thermometers. A discrepancy of a few mK is observed between these additional ROX sensors and the original OI installed sensors, with the difference between the QSHS and OI sensors on the MXC plate taken as the thermometry uncertainty in this paper.
    \par The tuning rod was held in a fixed position, at approximately 140 degrees. This angle was chosen such that the TM$_{\text{010}}$ mode's frequency was approximately in the centre of its available tuning range and to match planned axion search tests at the ADMX site. Here, the 0 degree point is defined with the tuning rod closest to the cavity wall (corresponding to the lowest TM$_{\text{010}}$ mode frequency) and 180 degrees defined with the tuning rod closest to the cavity center (highest TM$_{\text{010}}$ mode frequency). The rod was secured with a copper clamp-screw coupler mounted to the extruding sapphire axle at the bottom of the cavity. A copper flange was attached to one side of the axle clamp and then mounted to the base of the cavity itself, providing both mechanical support to hold the rod's vertical position and to act as a thermal sink. Further thermal sinking was implemented via a copper braid with one end attached to the axle clamp and the braid's other end attached to the OFC mount, this is shown in Figure \ref{fig:CavityBottView}. 
    \par
\begin{figure}[htbp]
    \centering
    \includegraphics[width=\linewidth]{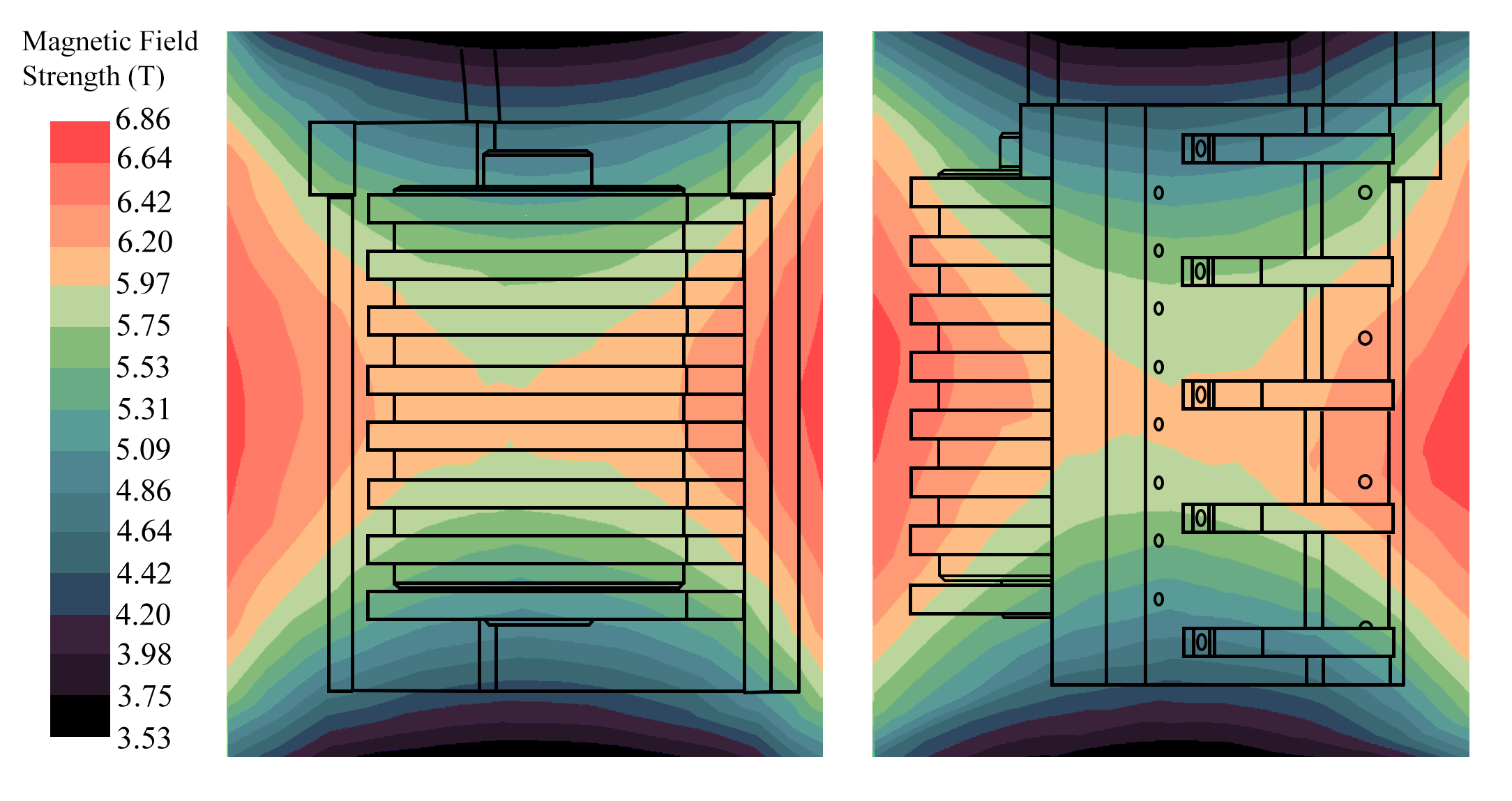}
    \caption{Field map of the integrated $\rm 6\,T$ superconducting magnet at full field with the position of the cavity mount and cavity indicated. Plot made from simulations in ANSYS Maxwell 3D \cite{ansyselectromagnetics} using current and coil information provided by Oxford Instruments \cite{oxinstproteox} under a non-disclosure agreement.}
    \label{fig:6TFieldMap}
    \end{figure}

 The magnet is a dry, superconducting solenoid with a maximum field of $\rm 6\,T$. Note that the magnet described in Reference \onlinecite{QSHSInstPaperArx} is an upgraded $8\, \rm T$ version of the one used here. The magnet uses a niobium-tin inner coil and two niobium titanium outer coils and is mounted in the DR via copper thermal shields which connect it to the second pulse tube (PT2) plate at $\rm 4\,K$. A field map of the $\rm 6\,T$ magnet is shown in Figure \ref{fig:6TFieldMap} with the cavity position indicated. The cavity is mounted off-center from the magnet's inner bore to allow the mounting block to also house static frequency cavities for other tests; it therefore experiences a non-uniform field. The field strength given here is calculated from ANSYS Maxwell 3D simulations \cite{ansyselectromagnetics} built on current and coil information provided by OI. The field strength shown in figures throughout is taken from these simulations as the mean field over the cavity volume with uncertainty the standard deviation, while critical fields in tables use the field evaluated at the relevant material surface (e.g. the cavity inner wall for $\mathrm{Al}$ and the tuning rod outer wall for $\mathrm{Nb}$ and $\mathrm{Nb_{3}Sn}$). A total of $\rm 120\,A$ of current is needed to ramp the magnet to full $6\,\rm T$ field.
    
    \subsection{Radio Frequency Instrumentation}
    To allow for full unloaded $Q$ calculations, both transmission and reflection scattering (S) parameter measurements are needed. The cavity and RF lines were set up such that all 4 S-parameters could be measured throughout the run. Two antennas were inserted into the cavity, both of these were kept at fixed insertion throughout tests and both were set to be undercoupled to the $\rm TM_{010}$ mode at room temperature (RT). In an axion search one of these ports has variable coupling and would be maintained at critical coupling throughout, this is designated Port 2 here, and the other as Port 1. In the absence of variable coupling both antennas were kept well into the undercoupled regime for these tests, for Port 2 this was done by identifying the critical coupling point manually as the point of maximal trough depth at the resonant frequency and then reducing insertion by approximately $2\,\rm mm$. Port 1 was kept at slightly lower insertion than Port 2, at its standard operating insertion of approximately $2\,\rm mm$. Both antennas were clamped with modified Staubli connectors \cite{Staubli} with internal BeCu fingerstock and held at fixed insertion with grub screws.
    
    \par
\begin{figure}[tbp]
    \centering
    \includegraphics[width=1.0\columnwidth]{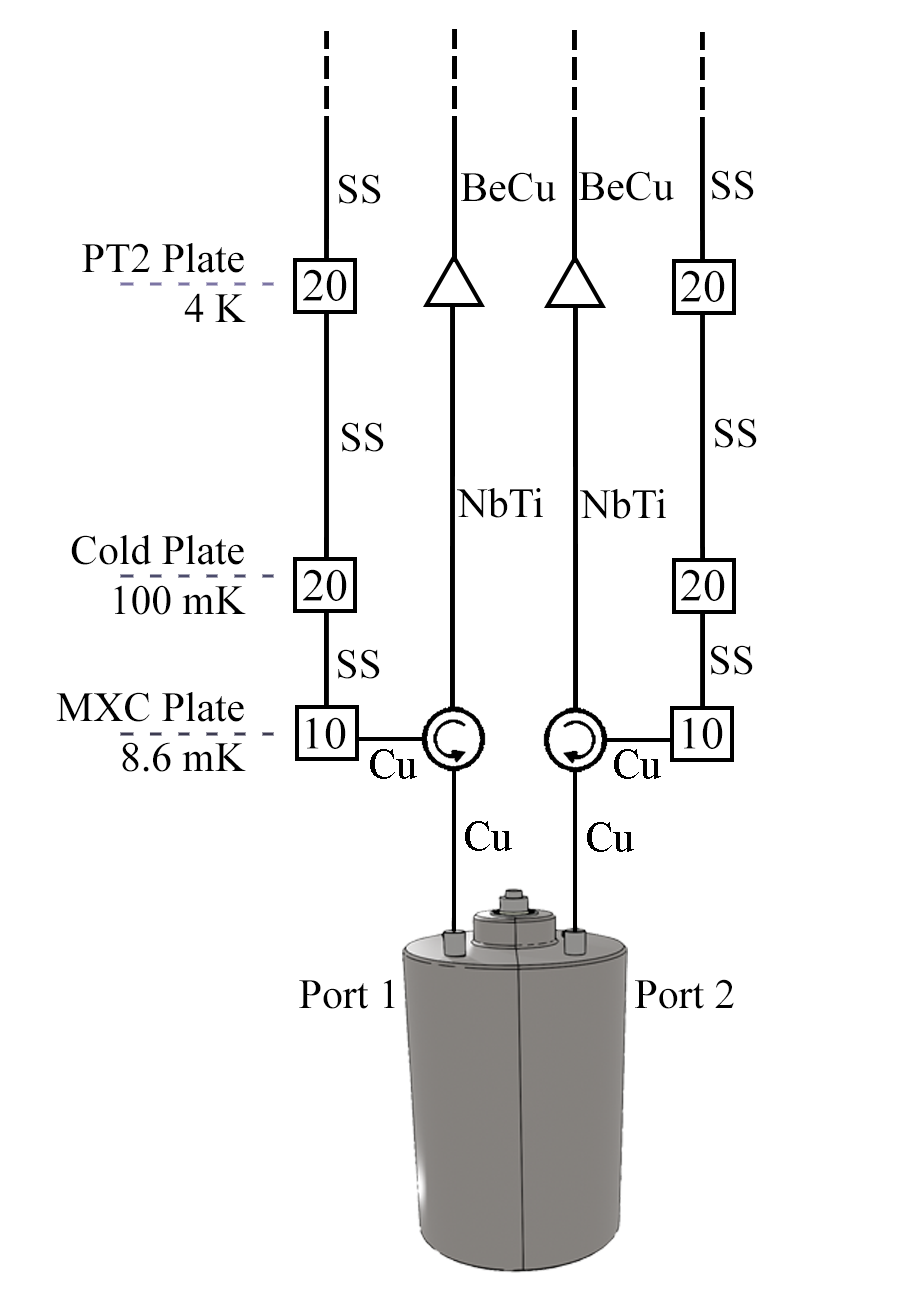}
    \caption{A circuit diagram of the radio frequency connection scheme in the dilution refrigerator insert. The circulators shown are Low Noise Factory LNF-CIC4\_8A circulators \cite{LowNoiseFactoryCirc}, and the amplifiers are both Low Noise Factory LNF-LNC4\_8G amplifiers \cite{LowNoiseFactoryCirc} with gain $40\, \rm dB$. Cables are all RG405 coaxial cables with inner and outer conductors made of the material given to the right of the line (or beneath for the circulator-input line lengths). During the run, modes were excited using the attenuated input lines (the two outer lines) and signals from the cavity detected using the superconducting signal lines (the two inner lines). RT switches allowed for all 4 scattering parameters to be measured. All RF lines are thermally sunk at each plate with gold-plated copper clamps. The initial cabling between the circulator and main dilution refrigerator RF cabling used NbTi for the signal lines and copper for the input lines (as indicated) with each of these $\approx 15\,\rm cm$ long, other cables run directly between plates or between the cavity ports and MXC plate with length as necessary for this distance and relief loops. The total one-way cabling length between the VNA and cavity is $\approx 11.5 \,\rm m$: with $\approx 2 \rm\, m$ of this within the insert.}
    \label{fig:RFChain}
    \end{figure}

 A diagram of the RF lines are shown in Figure \ref{fig:RFChain}. Both antennas fed into Low Noise Factory LNF-CIC4\_8A \cite{LowNoiseFactoryCirc} circulators, such that signals could be sent into either port through attenuated input lines and signals from the cavity were directed into NbTi superconducting signal lines for measurement. Both excitation and detection used an Anritsu ShockLine$^\text{TM}$ MS46522B series $20\,\rm GHz$ vector network analyzer (VNA) \cite{AnritsuVNA}. RF switches mounted above the DR were used to switch which lines the VNA ports connected to and allowed all 4 S-parameters to be measured.

\section{Data Acquisition}
\label{sec:dataAcq}
    \par The dataset taken consisted of VNA scans of 5 of the cavity's resonant modes; these modes and their frequencies are listed in Table \ref{tab:ModeFreqList}. The main figure of merit here is the improvement in the $\rm TM_{010}$ mode $Q$ as the mode relevant to axion searches. The $Q$ improvement to the other modes was also measured both for completeness and for their less-direct benefits to search runs, for example increased $Q$ in these non-target modes can result in less of the tuneable $\rm TM_{010}$ range being affected by mode crossings wherein the $\rm TM_{010}$ mode couples to other nearby modes and sees field distortion/hybridization and a reduced form factor and so sensitivity as a result.
    
    \begin{table}[htbp]
        \centering
        \begin{tabular}{c|c}
            Mode & Frequency (GHz) \\
            \hline
            \hline
            TM$_\text{010}$ & 5.27  \\
            TM$_\text{011}$ & 5.40 \\
            TE$_\text{111}$$^{\dagger}$ & 5.69 \\
            TE$_\text{110}$ & 5.82 \\
            TM$_\text{012}$ & 5.90 
        \end{tabular}
        \caption{The 5 modes monitored over the run and their resonant frequencies measured at cryogenic temperature and full-field. $^{\dagger}$ Splitting of the $\rm TE_{111}$ degenerate mode pair became resolvable for high $Q$ results, as noted in Section \ref{sec:analyres} and detailed in Appendix \ref{app:SpectandCoup}.}
        \label{tab:ModeFreqList}
    \end{table}

    \par The fridge reached a base temperature at the MXC plate of $\rm 10 \pm 4 \,mK$ and the cavity reached a base temperature of $\rm 70 \pm 10 \,mK$. Once at base temperature, the magnet was ramped to full field. Quenching or SC degradation of materials results in the release of their held currents and the sudden appearance of an effective resistive element, the resulting heating ($\mathcal{O}(\rm 0.7\,K)$) during these tests was high enough to require ramping be paused twice (once after the $\mathrm{Al}$ quench and again due to $\mathrm{Nb}$ degrading) to allow the temperature to return to the base point (as assessed by the MXC plate thermometry) and to keep temperatures well below any critical points. An initial ramp rate of $\rm 0.002\,T/min$ ($\rm 0.04\,A/min$) was used which was decreased to $\rm 0.001\,T/min$ ($\rm 0.02\,A/min$) after the first pause. 
    
    \par During both initial DR cooling and magnet ramping, measurements of the S-parameters were taken every 10 minutes with VNA input power $\rm 0 \,dBm$. Temperatures were logged every 20 seconds by the OI-calibrated thermometers and every 24 seconds for the additional thermometers. The magnet current was also logged every 20 seconds. The temperature readings used here in graphs and tables use the sensor mounted at the base of the cavity, described in Section \ref{sec:QSHSInst} and shown in Figure \ref{fig:CavityBottView}.
    
    \par The cavity transmission spectrum, measured as $S_{21}$ against frequency $\nu$, around each resonance is expected to follow a Lorentzian lineshape as

    \begin{equation}
    \label{eq:LorCav}
        S_{21}(\nu) = A \left[ \frac{1}{1+4\left( \frac{\nu-\nu_{0}}{\Gamma}\right)^{2}} \right]+\Delta S_{21}, 
    \end{equation}

    \noindent with $A$ the amplitude, $\nu_{0}$ the resonant frequency of the given mode, $\Gamma$ the mode's full-width-half-maximum bandwidth, and $\Delta S_{21}$ a shifting factor analogous to the insertion loss, modeled here as a linear $\Delta S_{21}=m\nu+c$. After fitting the power spectra over each resonance to the above, the loaded quality factor is found as

    \begin{equation}
        Q_{L} = \frac{\nu_{0}}{\Gamma}.
    \end{equation}

    The final unloaded $Q_{0}$ value, corrected for coupling is then found as

    \begin{equation}
        Q_{0} = Q_{L}(1+\beta_{1}+\beta_{2}),
    \end{equation}

    \noindent with the reflection coefficients found from the $S_{11}$ and $S_{22}$ reflection power spectra off each port and evaluated at the resonant frequency as

    \begin{equation} 
    \label{eq:beta1}
        \beta_{1} = \frac{1-|S_{11}|}{|S_{11}+S_{22}|} ,
    \end{equation}

    \noindent and 

    \begin{equation}
    \label{eq:beta2}
        \beta_{2} = \frac{1-|S_{22}|}{|S_{11}+S_{22}|}.
    \end{equation}

    \noindent Note the above uses the linear magnitude (not the dB value) of the S-parameters at the resonant frequency, corrected for the off-resonance baseline of the surrounding reflection spectra. Equations \ref{eq:beta1} and \ref{eq:beta2} assume both ports are in the same coupling regime (i.e. both either under or overcoupled), here they both remain undercoupled throughout the run as verified by phase fits to the complex data. A full description of the spectra analysis is given in Appendix \ref{app:SpectandCoup} with Appendix \ref{app:couplingphase} detailing the phase fits.
    
\section{Simulation}
\label{sec:sims}
    Modeling and simulation of the hybrid cavity was necessary to determine the expected quality factors of the Sidecar cavity under test. This was principally done by simulation of the cavity's sub-surface geometric factors in COMSOL and modeling the RF surface resistance of the surface materials.
    \subsection{Methodology of Hybrid SRF cavities}
    The quality factor of a multi-material cavity with $n$ materials, such that the total surface area $S=\sum_nS_n$, can be modeled as a summation of corresponding sub-surface quality factors, $Q_n$, for each material surface, $S_n$\cite{BraineMultiMode, BraineThesis}. Expressing this in terms of geometric factor, $G_n$, and the material surface resistance, $R_n$, is useful for separating the material properties from the geometry of the resonator. This can be written: 
    \begin{equation}
    \frac{1}{Q_{hybrid}}=\sum_n\frac{R_{n}}{G_{n}} ,
    \label{eqn:hybridN}
    \end{equation}
    where,
    \begin{equation}
    G_n=\frac{\mu\omega_0\int\vert\vec{H}|^2dV}{\int\vert\vec{H}|^2dS_n},
    \label{eqn:GsubFactor}
    \end{equation}
    where $\omega_0$ and $\vec{H}$ are the angular frequency and magnetic field of the resonant mode under study.
    \par In the case of the Cu-$\mathrm{Nb}_3\mathrm{Sn}$ ADMX Sidecar cavity system, one would simplify this into a copper cavity surface and $\mathrm{Nb}_3\mathrm{Sn}$ tuning rod surface. One can see that the ratio of this hybrid cavity's quality factor to an all copper version would reduce to:
    \begin{equation}
    \frac{Q_{hybrid}}{Q_{Cu}}=\frac{1+\frac{G_{\mathrm{Cu}}}{G_{\mathrm{Nb}_3\mathrm{Sn}}}}{1+\frac{R_{\mathrm{Nb}_3\mathrm{Sn}}}{R_{\mathrm{Cu}}} \frac{G_{\mathrm{Cu}}}{G_{\mathrm{Nb}_3\mathrm{Sn}}}}.
    \label{eqn:hybridkappaQratio}
    \end{equation}
    If one assumes ideal superconducting behavior, $R_{\mathrm{Nb}_3\mathrm{Sn}}\approx 0$, one can see that there is a limit to the improvement factor in $Q$ solely based on the geometric factors, albeit $1+\frac{G_{\mathrm{Cu}}}{G_{\mathrm{Nb}_3\mathrm{Sn}}}$. Based on the geometric factors shown at the end of this section for ADMX Sidecar, this is about $1.6$ for a tuning rod angle of 140 degrees (the angle these measurements were taken at), and varies from approximately $1.2-1.9$ over the entire tuning range.
    \par An additional characterization tool for multi-material cavities is multi-mode decomposition. Whereas $G_n$ above refers to the $G$ factor of the $n$th surface for a single resonant mode, this can be expanded to $G_{mn}$ where $m$ refers to the $m$th resonant mode; this forms a matrix of geometric factors. By making measurements of multiple cavity modes' quality factor, one can, through a matrix inversion, calculate the "in-situ" resistance of each cavity surface, as long as they can precisely simulate $G_{mn}$. The principal limitation of this process is having modes whose geometric factors are statistically distinct for the surfaces of interest; if they are not, the uncertainty in resistance can be too high to distinguish the difference in surface resistance. The details of this process are outlined in References \onlinecite{BraineMultiMode, BraineThesis}.
    \subsection{Modeling Material Surface Resistance}
    \par In order to calculate the expected hybrid cavity $Q$, as shown in Equations \ref{eqn:hybridkappaQratio} and \ref{eqn:hybridN}, one must know the surface resistance of the materials in use. Since the cavity will contain normal and superconducting states through its various transitions in the cooling process, knowledge of resistance in both states is useful. 
    \par In the case of normal conductors, we can follow a semi-classical formulism to estimate resistance. Classically, the resistance can be calculated from the conductivity of the material, $\sigma$, or its classical skin depth, $\delta$: 
    \begin{equation}
    R=\sqrt{\frac{\mu_c \omega}{2\sigma}}=\frac{\mu_c\delta \omega}{2},
    \label{eqn:Rclassic}
    \end{equation}
    where $\mu_c$ is the material permeability and $\omega$ is the angular frequency of the resonant mode.
    At cryogenic temperatures, however, the conductivity plateaus as the mean free path of electrons in the normal metal exceed the classical skin depth; This can be thought of as an "anomalous" skin depth regime \cite{BoutanThesis,HotzThesis}, which can be written \cite{HagmannThesis}: 
    \begin{equation}
    \delta_{anom}=\left(\frac{\sqrt{3} c^2 m_e v_F}{8 \pi^2 \omega_0 n e^2}\right)^{1/3},
    \label{eqn:danom}
    \end{equation}
    where $m_e$ is the electron mass, $e$ is the electron charge, $v_F$ is the fermi velocity of the material, $n$ is the conduction electron density. One can use the parameters, $v_F= 1.57 \times 10^8\,{\rm cm/s}$ and $n=8.45 \times 10^{22}\,{\rm cm^{-3}}$ for copper and $v_F= 2.02 \times 10^8\,{\rm cm/s}$ and $n=6.00 \times 10^{22}\,{\rm cm^{-3}}$ for aluminum to estimate this anomalous skin depth to be\cite{Kittel2004} :
    \begin{equation}
    \delta_{anom,Cu}=(2.84\times 10^{-4}\,{\rm m\cdot Hz^{1/3}})\cdot \nu^{-1/3},
    \label{eqn:danomCU}
    \end{equation} 
    \begin{equation}
    \delta_{anom,Al}=(3.46\times 10^{-4}\,{\rm m\cdot Hz^{1/3}})\cdot \nu^{-1/3}.
    \label{eqn:danomAl}
    \end{equation} 
    Assuming a resonant frequency of $5.2\,\mathrm{GHz}$, the surface resistance of pure copper and aluminum would be, $R_{\mathrm{Cu}}=3.38\,\mathrm{m \Omega}$ and $R_{\mathrm{Al}}=4.01\,\mathrm{m \Omega}$.
    \par To account for the 6061 aluminum alloy, empty Sidecar cavity measurements were made that were published in Reference \onlinecite{Esmenda_2025}. The normal cryogenic resistance of the empty cavity TM$_\text{010}$ mode was measured to be $R_{\mathrm{6061Al}}(2.5\,\mathrm{GHz})=10.3\pm0.2\,\mathrm{m \Omega}$. Following the same frequency relation as Equation \ref{eqn:Rclassic}, the resistance of the 6061 aluminum can be extrapolated to be $R_{\mathrm{6061Al}}(5.2\,\mathrm{GHz})=16.8\pm0.3\,\mathrm{m \Omega}$ for the loaded cavity.
    \par The anomalous skin depth of niobium and $\mathrm{Nb}_3\mathrm{Sn}$ in their cryogenic normal states are not well documented. This may be because pure niobium's residual-resistivity ratio (RRR) varies wildly (3-90000) based on its specific composition\cite{Niobiumprop2,Niobiumprop3}. For pure niobium, one can find in Reference \onlinecite{PhysRev.181.1127} the low temperature normal resistivity to be approximately $1.6\times 10^{-7} \,\mathrm{\Omega \cdot m}$, which would correspond to a surface resistance of $R_{\mathrm{Nb,\,Norm}}(5.2 \, \mathrm{GHz})\approx 81\, \mathrm{m\Omega}$. $\mathrm{Nb}_3\mathrm{Sn}$ resistance can be found at $20\,\mathrm{K}$ to be $R_{\mathrm{Nb}_3\mathrm{Sn},\,\mathrm{Norm}}(15 \, \mathrm{GHz})=94.6\, \mathrm{m\Omega}$ in Reference \onlinecite{Alimenti_2021}; one can correct for frequency using Equation \ref{eqn:Rclassic} giving a surface resistance $R_{\mathrm{Nb}_3\mathrm{Sn},\,\mathrm{Norm}}(5.2 \, \mathrm{GHz})=46.7\, \mathrm{m\Omega}$, which is nearly 14 times that of copper. In short, although the specific resistance varies significantly with composition, niobium and its compounds in the normal state have a significantly higher RF resistance than copper and aluminum. This means that the hybrid system in the normal state will have a low $Q$ dominated by the very resistive niobium component.
    \par It is also relevant to model the surface resistance of the superconducting states. As alluded to by Equation \ref{eqn:hybridkappaQratio}, one can assume $R_{SC}\approx 0$ for ideal superconductors within a hybrid cavity and get a finite limit for $Q$. However, if the entire cavity is superconducting, such as the Aluminum-$\mathrm{Nb}_3\mathrm{Sn}$ cavity at $25\,\mathrm{mK}$, this approximation produces an infinite result.
    \par The empty cavity measurements in Reference \onlinecite{Esmenda_2025} also provide us with an estimate and model for the superconducting aluminum losses. The resistance of the type I superconducting aluminum surface can be modeled with 3 components:
    \begin{equation}
    R_{Type\ I, \,SC}=R_{BCS}+R_{TLS}+R_{residual}.
    \label{eqn:RSCTypeI}
    \end{equation} 
    The BCS Resistance can be written:
    \begin{equation}
        R_{BCS}\propto \lambda^3l\omega^2 e^{-1.76T_c/T},
        \label{BCSResistance}
    \end{equation}
    where $\lambda$ is the penetration depth and $l$ is the mean free path of unpaired electrons. The TLS losses stem from non-aluminum impurities in the alloy, or lattice defects, that form a two energy level system. This resistance can be written:
    \begin{equation}
        R_{TLS}\propto \beta \tanh(\frac{\hbar \omega}{2 k_B T}),
        \label{TLSResistance}
    \end{equation}
    where $\beta$ is a fitting factor. The residual resistance, $R_{residual}$, is a constant independent of temperature due to impurities in the superconductor. One can see that the BCS resistance will be at its maximum near the critical temperature ($T_c\approx1.2\,\mathrm{K}$ for aluminum), and as it is cooled, it will reach a plateau based on the residual. This plateaued resistance was measured for the empty cavity $\mathrm{TM_{010}}$ mode to be $R_{6061\mathrm{Al},\,\mathrm{SC}}(2.5 \, \mathrm{GHz})=12.2\pm0.5\, \mathrm{\mu\Omega}$. The residual resistance of the $\mathrm{Nb}_3\mathrm{Sn}$ tuning rod could not be measured in this way because it would require an empty cavity entirely made of the $\mathrm{Nb}_3\mathrm{Sn}$ with the same composition as the tuning rod.
    \par Type II superconductor's resistivity is also governed by vortex motion due to magnetic fields; one can use the simplified Gittleman-Rosemblum model \cite{GRfluxonmodel}, which assumes the gross motion is an average effect of a single fluxon\cite{PompeoSilva}. One finds that the resistance of this motion grows significantly with increasing magnetic field as the material approaches its second critical field, $B_{c2}$. Following this model, the low-field resistance can be approximated by Taylor expansion about $B=0$, where $B$ is the applied magnetic field:
    \begin{equation}
        R_{vm}(B) \approx \sum_{n=0}^{N}{R_n \left( \frac{B}{B_{c2}}\right)^{2n+1}}.
    \end{equation}
    Solving for the first order term in the Taylor series gives a resistance constant
    \begin{equation}
        R_0= \frac{\rho_0}{2\lambda}\frac{1}{(\frac{\omega_p}{\omega})^2+1},
        \label{eqn:SRFR_s}
    \end{equation}
    The value of $\rho_0$ is a constant set by the manufacturing, depending on the observed grain size, with a smaller grain size translating to a higher resistivity \cite{Posen_2017}; in the case of $\mathrm{Nb}_3\mathrm{Sn}$, $\rho_0=0.71-9.7\times 10^{-7} \,{\rm \Omega\cdot m}$ according to the manufacturers of the tuning rod at Fermilab. For $\mathrm{Nb}_3\mathrm{Sn}$, $B_{c2}\approx30\,\mathrm{T}$, and the de-pinning frequency $\omega_p\approx 6.5\,\mathrm{GHz} $\cite{Keckert_2019}. For pure niobium,  $B_{c2}\approx0.3\,\mathrm{T}$, and $\nu_p\approx 673\,\mathrm{MHz} $\cite{NbPinningFreq,PhysRev.139.A1515,PhysRev.139.A1515,PhysRevAccelBeams.23.023102}. These parameters suggest any exposed pure niobium substrate on the tuning rod, as is hypothesized in our experimental results later, will contribute a significant surface resistance dominated by vortex motion, even near the transition from the Meissner to mixed state. $\mathrm{Nb}_3\mathrm{Sn}$, on the other hand, will most likely be dominated by residual resistance due to impurities in the coating in its initial transition to a mixed state. This is why we highlight the low-field approximation.
    \par This Taylor approximation breaks down for high fields, however, and can't be used accurately at $\mathcal{O}(1\,\mathrm{T})$. Additionally, this simplified model does not account for the dependency of the angle of the external field with respect to the normal vector of the material surface; This level of modeling constitutes a publication in itself. The low field approximation was sufficient for interpretation of our experimental results.
    \subsection{Simulation of ADMX Sidecar}
    \par The geometric factors for the relevant modes were simulated using the COMSOL Multiphysics\textregistered{}  emw eigenmode software package\cite{COMSOL}, and are listed in Table \ref{tab:sidecarGfactors}. The CAD model of the cavity was imported into COMSOL and exterior features, such as tapped holes, were removed for simplicity and faster calculation. The correct position of the tuning rod was determined by finding the angle that corresponded to the TM$_\text{010}$ mode frequency. The higher modes were then determined by matching the simulated frequency to observed frequency in testing. The simulated mode fields were then integrated to arrive at a geometric factor value for both the exterior cavity, the tuning rod, and the copper tuning rod axle tabs.
    \par Other smaller interior surfaces, such as the tuning rod axles, were determined to be two orders of magnitude higher, $\mathcal{O}(10^4\,\Omega)$, than the rod and cavity, $\mathcal{O}(10^2\,\Omega)$, and therefore did not have a significant impact on the total geometric factor. The silver-plated mounting screws of the axle tabs, and were modeled as a flat copper surface instead; the conductivity difference of silver and copper was deemed a negligible difference on the total sub-surface quality factor of the tab surface. However, the $R_{\rm s}/G_{\rm s}$ could still be significant for more resistive sapphire surfaces; because this sapphire has a loss tangent of $\mathcal{O}(10^{-4})$ this would not be the case for the axles.
    \par The error in the geometric factor was estimated by performing a parametric sweep over the possible variations in the tuning rod and cavity dimensions according to the machine tolerances ($\pm0.005"$). To achieve this, the CAD geometry of the rod and cavity were individually given an isotropic scaling by a factor equivalent to this length tolerance, which was approximately $\approx\pm0.1\%$. The geometric factors for a given surface were averaged over the values produced by each variation in scaling, and the error was taken to be the standard deviation of this set. 
    \par It was the intention to use these geometric factors to perform a multi-mode decomposition in order to solve for the surface resistances of the cavity and tuning rod during our measurements, as is done in Reference \onlinecite{BraineMultiMode}. This process requires taking the inverse of a matrix made of geometric factor weights, which necessarily must have a statistically non-zero determinant to produce physical results. This determinant is near-zero when modes have near-identical geometric factors. Upon calculating the determinant of these matrices for the Sidecar system using the geometric factors in Table \ref{tab:sidecarGfactors}, it was realized no combination of modes met this condition, and therefore a multi-mode decomposition was not possible for this tuning rod position. This position was chosen because its corresponding $\rm TM_{010}$ frequency was within a frequency region that improved axion-photon limits could be made in a future search. A full analysis of optimal tuning rod position for multi-mode decomposition, although possible, was found to extend beyond the practical limits of the computational resources and time available. Additionally, such an analysis may show there is no tuning rod arrangement where a surface decomposition is possible with any combination of the modes under study; if so, one could pursue higher order modes for decomposition study, but this becomes experimentally harder to measure due to mode crowding. Instead, we rely on Equation \ref{eqn:hybridkappaQratio} to make inferences on the resistance of the superconducting tuning tod with respect to aluminum and copper in the analysis of our results.
\begin{table} [ht]
    \centering
    \begin{tabular}{c| c| c |c | c } 
    Mode & Exterior Cavity & Tuning Rod & Axle Tabs & Total \\ [0.5ex] 
        \hline
        \hline
        TM$_\text{010}$ & $237\pm42$ & $327 \pm 58$ & $1355 \pm 339$ & $125 \pm 40$ \\
        TM$_\text{011}$ & $197\pm36$ & $294 \pm 51$ & $1223 \pm 371$ & $108 \pm 40$ \\
        TE$_\text{111}$ & $280\pm6$ & $355 \pm 8$ & $5898 \pm 834$ & $153 \pm 22$ \\
        TE$_\text{110}$ & $270\pm25$ & $411 \pm 35$ & $2116 \pm 468$ & $151 \pm 38$ \\
        TM$_\text{012}$ & $381\pm49$ & $528 \pm 70$ & $5415 \pm 1553$ & $213 \pm 73$ \\ 
        \end{tabular}
        \caption{The relevant geometric factors in Ohms of the Sidecar cavity for a tuning rod angle of 140 degrees from the closest approach to the cavity wall. This angle corresponds to the TM$_\text{010}$ mode frequency that the Sidecar cavity was tuned to during testing. The rod geometric factor represents the surface not including its axle tab inserts. Surfaces other than the rod, axles tabs, and exterior cavity were found to be $\rm >10^4\,\rm \Omega$, and thus were insignificant to the total.}
        \label{tab:sidecarGfactors}
        \end{table}

\begin{table} [ht]
    \centering
    \begin{tabular}{ c |c } 
        Cavity State& Predicted $Q_0$ \\ [0.5ex] 
        \hline
        \hline
        All $\mathrm{Cu}$& $39,000\pm7,500$ \\
        All $\mathrm{Al}$ & $8,000\pm1,000$ \\
        Normal $\mathrm{Nb}_3\mathrm{Sn}+ \mathrm{Al}$ & $4,600\pm600$ \\
        Superconducting $\mathrm{Nb}_3\mathrm{Sn} $ + Normal $\mathrm{Al}$& $13,600\pm2,400$ \\
        Superconducting $\mathrm{Nb}_3\mathrm{Sn} $ + Normal $\mathrm{Cu}$& $59,500\pm9,400$ \\
        Superconducting $\mathrm{Nb}_3\mathrm{Sn}+ \mathrm{Al}$ & $3.8\pm1.0\times10^5$ \\

        \end{tabular}
        \caption{The estimated zero field unloaded quality factors for the $\mathrm{TM_{010}}$ mode. This was calculated for the normal state, the state where only $\mathrm{Nb}_3\mathrm{Sn}$ is superconducting, and the state where both $\mathrm{Nb}_3\mathrm{Sn}$ and Aluminum superconduct. Additionally, the prediction for an all copper cavity and rod and all normal 6061 aluminum cavity and rod are included for reference, as well as an estimate for a copper cavity with a $\mathrm{Nb}_3\mathrm{Sn}$ rod in the superconducting state. All estimates assume the tuning rod axle tabs are made of copper. The error is calculated from uncertainty in the G factor simulations, as well as uncertainty in the resistivity for the aluminum contributions as they are well-defined from experimental data in Ref. \onlinecite{Esmenda_2025}. Uncertainty in copper and niobium components' resistivity was not well-known, and not included, therefore, it must be noted that this is an underestimate of the uncertainty.}
        \label{tab:Q_est}
        \end{table}
    \par An expected $Q$ value at zero field can be calculated using the expected surface resistance of the various materials. Table \ref{tab:Q_est} shows these values for the cavity $\mathrm{TM_{010}}$ mode in its normal state, the state where only $\mathrm{Nb}_3\mathrm{Sn}$ is superconducting, and the state where both $\mathrm{Nb}_3\mathrm{Sn}$ rod and Aluminum cavity are superconducting. Of particular note is that the expected $Q$ for the state where only the rod is superconducting is actually less than for an all copper cavity, which is a result of the increased resistance of the normal 6061 aluminum based on empty cavity measurements. 
    \par The fully superconducting state is expected to be an overestimate because it does not account for the residual resistance of the $\mathrm{Nb}_3\mathrm{Sn}$ coating. Nonetheless, this quality factor is dominated by the copper axle tabs because of its much higher resistance; even if the coating had a resistance 10 times that of the superconducting aluminum, this would only reduce the $Q$ by 55\%. The copper resistance is determined by the anomalous skin depth model described, which is an ideal, and may also be higher for the particular alloy of the axles tabs, similar to 6061 aluminum; this could significantly reduce the estimated $Q$. A factor of two increase in copper resistance lowers the fully superconducting $Q$ to 192,000. 

\section{Results}
\label{sec:analyres}

\subsection{Cool-down results}
\label{sec:CoolRes}

    \par
\begin{figure}
        \centering
        \includegraphics[width=\linewidth]{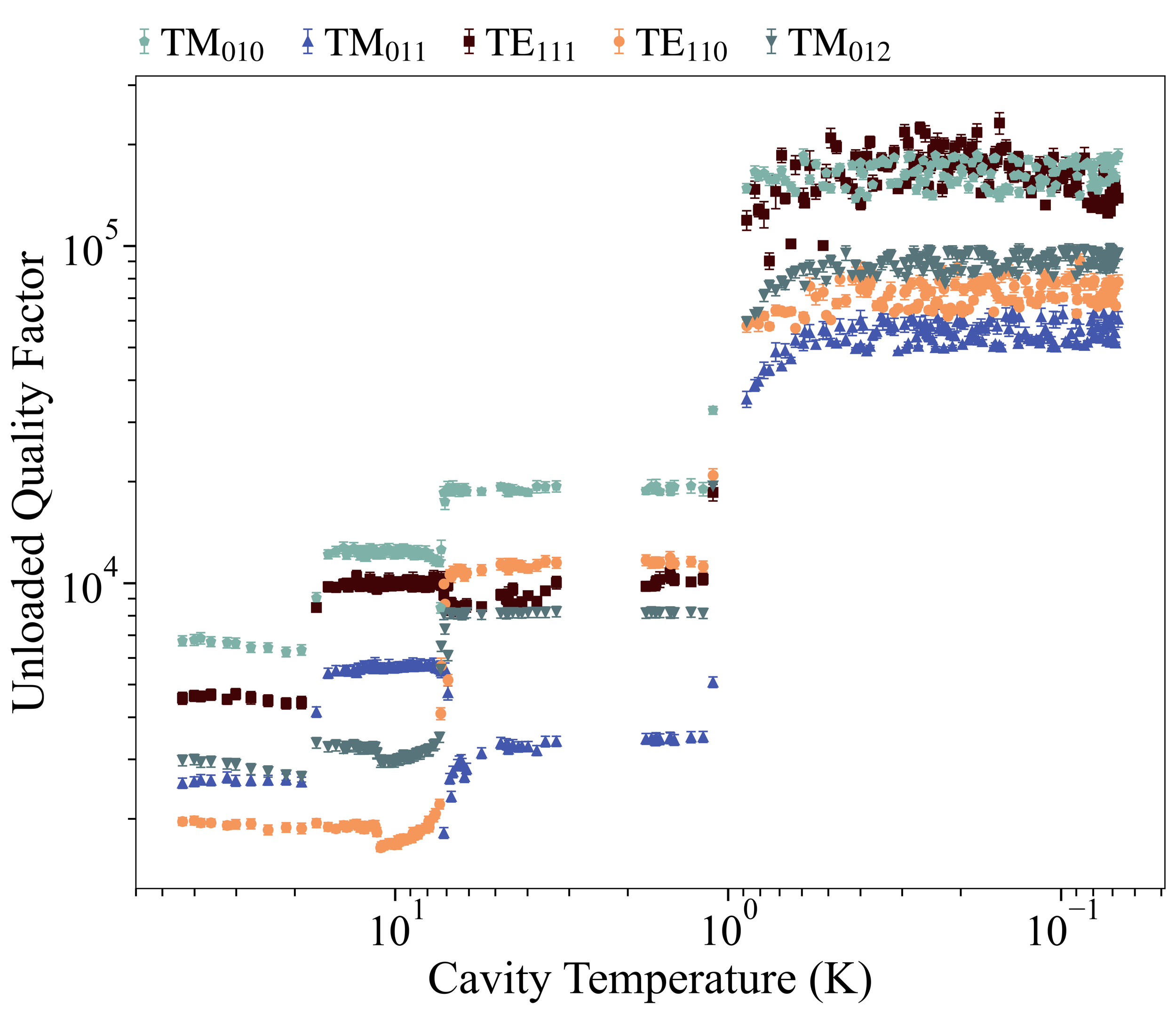}
        \caption{The unloaded quality factors of a number of cavity resonances (Table \ref{tab:ModeFreqList}) by temperature during dilution refrigerator cooling. The $Q_{0}$ uncertainties are based on the change in result when considering different data ranges about the resonance as described in Appendix \ref{app:SpectandCoup}. The apparent gap in the data range between 2 and $3\, \rm K$ is due to the temperature dropping sharply past this range (see Figure \ref{fig:CoolingQsByTime}).}
        \label{fig:CoolingQsByTemp}
    \end{figure}

 The $Q$s of the five resonant modes with temperature during DR cooling is shown in Figure \ref{fig:CoolingQsByTemp}. Three distinct superconducting transitions were visible, corresponding to the critical temperatures of $\mathrm{Nb}_3\mathrm{Sn}$, $\mathrm{Nb}$, and $\mathrm{Al}$, chronologically. Each resulted in a notable change in the $Q_0$ of the cavity resonances, with these transitions occurring over the temperature ranges given in Table \ref{tab:CoolingCritTemps}. 
    
    \par
\begin{figure}
        \centering
        \includegraphics[width=\linewidth]{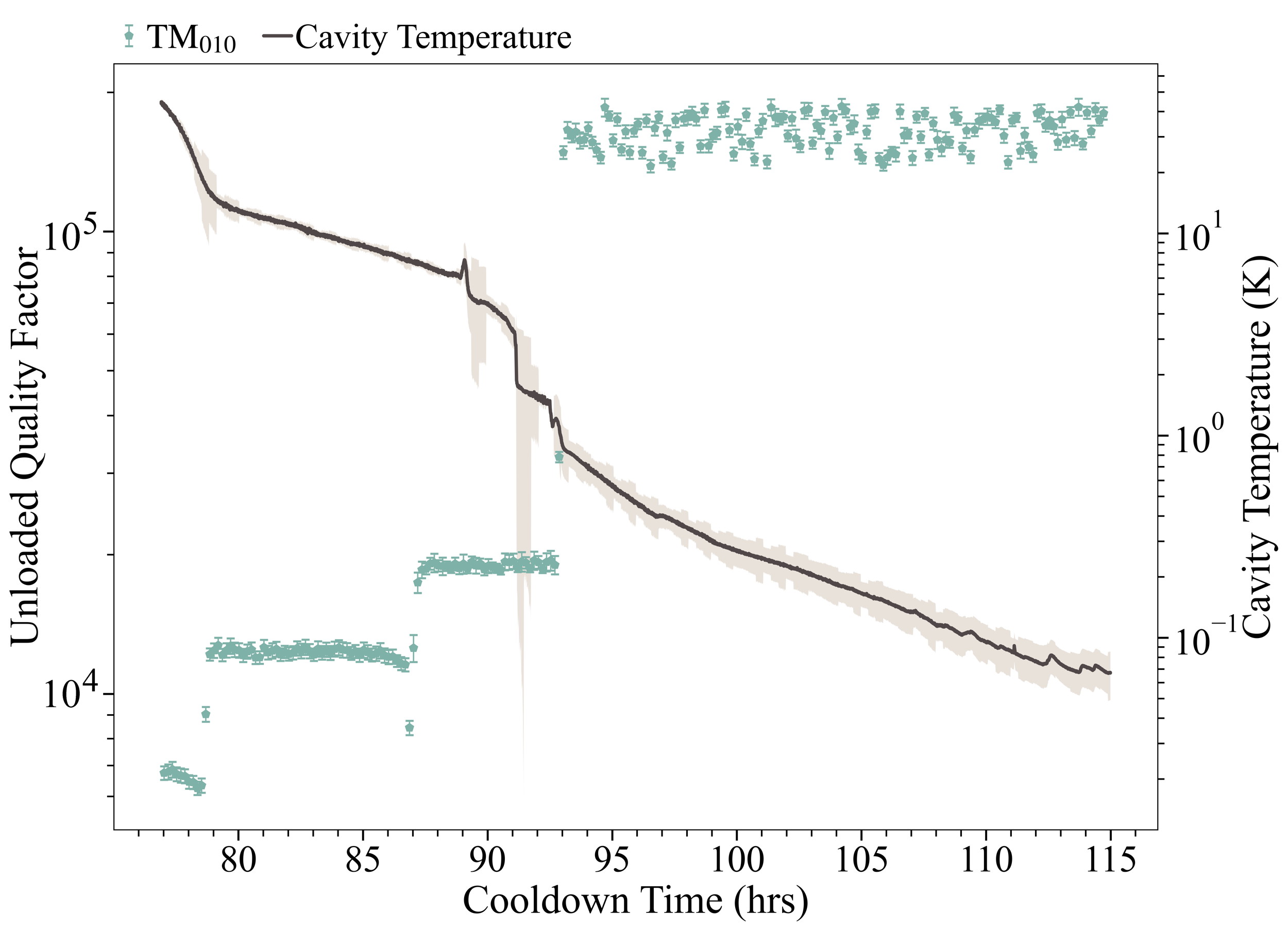}
        \caption{The cavity temperature and $\rm TM_{010}$ mode by time during dilution refrigerator cooling. The uncertainties indicated on the temperature values are taken as the discrepancy between the Oxford Instruments-installed temperature sensors and the QSHS-installed sensors on the mixing chamber plate. The regions of high apparent temperature uncertainty are due to slight delays between the OI-installed and QSHS-installed thermometers on the MXC plate which become apparent after sharper spikes or drops in temperature, the corresponding uncertainty due to lag between the cavity thermometer and tuning rod may be higher than that indicated here.}
        \label{fig:CoolingQsByTime}
    \end{figure}

 Transitions generally begin slightly below the expected critical temperature. The thermometer sits at the cavity base, while the tuning rod is somewhat thermally isolated by its sapphire axles and so cools more slowly than the directly thermally linked cavity. The inner cavity and rod therefore lag the thermometer in temperature, reaching critical temperatures only once the thermometer has already cooled below it. The exception is the $\mathrm{Nb_3Sn}$ transition, which begins slightly above its expected critical temperature, likely due to variation in said critical temperature with the exact stoichiometry and thickness of the coating. The time taken for the rod and cavity to fully equilibrate also accounts for the spread in temperature over which each transition is observed, as not all of a given component reaches its critical temperature at once. The cavity cooling curve is shown in Figure \ref{fig:CoolingQsByTime} along with a plot of the target $\rm TM_{010}$ mode $Q_0$ specifically.
    
    \begin{table}[ht]
        \centering
        \begin{tabular}{c|c|c|c}
            SC Material & $T_{c}$ (K) & Duration (hrs) & Expected $T_{c}$ (K) \\
            \hline
            \hline
            $\mathrm{Nb}_3\mathrm{Sn}$ & 19.2 -- 15.9 & 0.3 & 18 \\
            $\mathrm{Nb}$ & 7.38 -- 6.68 & 1.1 & 9.2 \\
            $\mathrm{Al}$ & 1.19 -- 0.567 & 2.3 & 1.2
        \end{tabular}
        \caption{The observed transition temperatures $T_{c}$ by material for the cavity during dilution refrigerator cooling, here the $\mathrm{Al}$ makes up the cavity, the $\mathrm{Nb}_3\mathrm{Sn}$ the coating on the tuning rod, and the $\mathrm{Nb}$ the bulk substrate of the tuning rod. The range given for $T_{c}$ reflects the temperatures at which transitional behavior began and ended, with duration being the time elapsed between these points. The cavity and internal tuning rod are suspended as shown in Figure \ref{fig:QSHSDRDiag} below the mixing chamber plate and will cool non-uniformly resulting in more gradual transitions at the critical temperatures. Literature values for critical temperatures are taken from Reference \onlinecite{CritFieldofSCAl} for $\mathrm{Al}$ and Reference \onlinecite{Nb3SnForCavs} otherwise.}
        \label{tab:CoolingCritTemps}
    \end{table}

    \par
\begin{figure*}
        \centering
        \includegraphics[width=\linewidth]{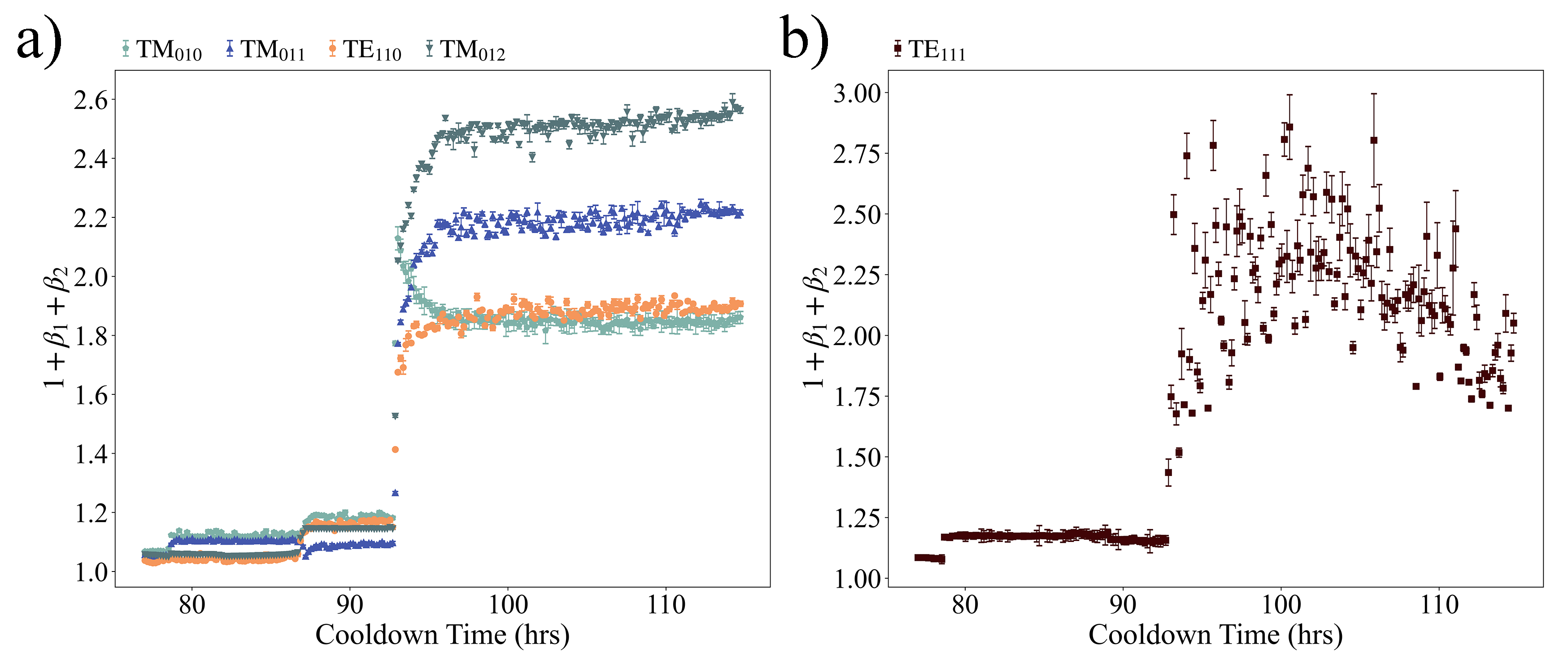}
        \caption{The coupling correction factor by time during dilution refrigerator cooling with (b) the $\rm TE_{111}$ mode results and (a) all other modes measured. The $\rm TE_{111}$ mode is plotted separately here for clarity as it saw enough degeneracy lifting between the two degenerate TE mode field orientations of these to be separately resolvable at the high $Q$s post-$\mathrm{Al}$ transition, this introduces effective mode-crossing-like behavior between the two modes and thereby significant scatter and more complex coupling evolution.
        These values are used for calculating the unloaded $Q$s shown in Figure \ref{fig:CoolingQsByTemp} and Table \ref{tab:CoolingQs}. Calculation details are given in Appendix \ref{app:SpectandCoup}.}
        \label{fig:CoolingCouplings}
    \end{figure*}

 Figure \ref{fig:CoolingCouplings} shows the coupling correction factor with time during cooling. The overall trend seeing coupling increase with each transition is consistent with the definition of the coupling parameters as $\beta_{i}=Q/Q_{C,i}$ with $i$ indexing the ports, $Q$ being the unloaded $Q$, and $Q_C$ representing the coupling quality factor. As $Q_{C,i}$ is set predominantly by the fixed antenna insertion depth and should be largely constant over the run, the increase in $Q$ therefore drives a corresponding increase in $\beta$. The decrease in $\beta$ for the $\rm TM_{011}$ mode after the $\mathrm{Nb}$ transition is likewise consistent with the degraded $Q$. The exception to this is the coupling of the $\rm TM_{010}$ mode after the $\mathrm{Al}$ transition, which sees an initial jump as expected followed by a smooth decline in $1+\beta_{1}+\beta_{2}$ which then plateaus at a lower value. As this relaxation is not seen in the $\rm TM_{010}$ mode's $Q$ (Figure \ref{fig:CoolingQsByTime}) it cannot be due to some transient change in the bulk surface resistance of the $\mathrm{Al}$ occurring post transition and identifying an exact cause as e.g. changes in current distribution post-transition is difficult. A similar effect is seen in the coupling of the $\rm TE_{111}$ mode, however due to its much larger scatter the overall trend and plateau is harder to ascertain, it was observed that the degeneracy of this TE mode was enough to make the two peaks separately resolvable  in the high $Q$ region where all materials are superconducting. This degeneracy lifting is discussed in full in Appendix \ref{app:SpectandCoup}, but results in mode-crossing like behavior of the $\rm TE_{111}$ mode specifically where the exact degeneracy and coupling between the two split modes at high $Q$ and their merging into a single broader feature at lower $Q$ introduces significant noise to the data and scatter in both the coupling and $Q$ measurements. For both modes, however, this evolution in $\beta$ is a transient effect and settles after a short period of time, it does not therefore have any significance for the results herein or the applicability of the cavity to haloscope searches.
    
    \par The effect of the pure niobium transition was unexpected as the $\mathrm{Nb}_3\mathrm{Sn}$ coating, being in the bulk limit of thickness ($\rm 3 \,\mu m$), was expected to effectively shield the effects of the niobium substrate. This transition could imply either the presence of imperfections (scratches, thinner regions, etc.) in the tuning rod coating or a higher amount of flux penetration between the two layers than expected. This did not prevent the coating retaining some SC behavior as reflected in sustained $Q$ increase at high field (Section \ref{sec:RampRes}), however, it may still represent an issue for this coating process being used for future haloscopes. 
    \par If the tuning rod has regions where defects in the coating leave the $\mathrm{Nb}$ bulk either fully exposed or are such that the $\mathrm{Nb}_3\mathrm{Sn}$ coating enters the thin film regime (and not the bulk as expected), this creates difficult-to-quantify uncertainties in geometric factors simulations. These uncertainties are due to variation in the effective surface resistance of the rod as well as potential interactions of the relevant fields with the defects due to e.g. concentration of vortices from the $\mathrm{Nb}_3\mathrm{Sn}$ in its mixed state about these defects. To model these reliably then, it would be necessary to know the exact location and extent of scratches or thinner regions and even then would likely increase the required computation time for these simulations and introduce excess uncertainty to them.
    \par Particularly of note, imperfections on the end cap regions, would most likely have a disproportional effect on the performance, because of the field concentration in these regions, versus imperfection on the rod walls, making even a small fraction of exposure significant. Therefore, for this coating to be useful in future haloscopes, it is likely multiple coatings, better handling procedures, or improvements to the initial coating method may be needed. However, the accelerator community has achieved near perfect coverage in their $\mathrm{Nb}_3\mathrm{Sn}$ cavities with this process, so it is more likely that improper handling is responsible in this case. Careful handling will be especially important when rod motion is introduced, increasing the likelihood of rubbing between the rod end cap and cavity that could produce scratches. It is still interesting, however, to see if the benefit in $Q$ is retained at full field to assess the viability of the coating overall.
    \par Rather than coating imperfections, it is possible this observed transition is due to the dispelling of residual flux in the $\mathrm{Nb}_3\mathrm{Sn}$ coating by means of the Meissner effect of the niobium substrate when its goes superconducting. The $\mathrm{Nb}_3\mathrm{Sn}$, due to its low demagnetization factor, could be prone to retaining residual flux after its Meissner transition due to a significant background magnetic field. This could be due to residual current in the magnet when not operating as well as other external sources, such as the earth's magnetic field. If so, the bulk niobium substrate, during its transition to the superconducting state, could dispel this residual background flux from the coating, causing an increase in the quality factor.
    
    \par The overall gains in $Q$ through each transition are given in Table \ref{tab:CoolingQs}. The $\mathrm{Nb}_3\mathrm{Sn}$ coating transition sees a factor 1.86 improvement to the target $\text{TM}_{010}$ mode $Q$. Using Equation \ref{eqn:hybridkappaQratio} and the expected all aluminum cavity and rod $Q$ from Table \ref{tab:Q_est}, with a slight correction to account for the axle tabs, one can estimate that $R_{rod}=2.68\,\rm m\Omega$, 16\% that of aluminum, and 79\% that of copper at cryogenic temperatures. 

    \begin{table}[ht]
        \centering
        \begin{tabular}{c|c|c|c|c}
            $T_{\text{cav}}$ (K) & SC Materials & Mode & $Q$ & $Q/Q_{i}$ \\
            \hline
            \hline
             &  & TM$_\text{010}$ & $6590 \pm 80$ & 1 \\
             &  & TM$_\text{011}$ & $2600 \pm 30$ & 1\\
            > 19.2 & None & TE$_\text{111}$ & $4570 \pm 60$ & 1\\
             &  & TE$_\text{110}$ & $1930 \pm 20$ & 1 \\
             &  & TM$_\text{012}$ & $2880 \pm 40$ & 1 \\             
            \hline
             &  & TM$_\text{010}$ & $12260 \pm 60$ & $1.86 \pm 0.01$ \\
             &  & TM$_\text{011}$ & $5630 \pm 30$ & $2.17 \pm 0.01$ \\
            15.9 -- 7.38 & $\mathrm{Nb}_3\mathrm{Sn}$ & TE$_\text{111}$  & $9900 \pm 50$ & $2.17 \pm 0.02$\\
             &  & TE$_\text{110}$ & $1820 \pm 40$ & $0.94 \pm 0.01$ \\
             &  & TM$_\text{012}$ & $3140 \pm 20$ & $1.090 \pm 0.008$  \\
            \hline
             &  & TM$_\text{010}$ & $18900 \pm 100$ & $2.87 \pm 0.02$\\
             &  & TM$_\text{011}$ & $3230 \pm 80$ & $1.24 \pm 0.02$ \\
            6.68 -- 1.19 & $\mathrm{Nb}_3\mathrm{Sn}$, $\mathrm{Nb}$ & TE$_\text{111}$ & $9000 \pm 300$ & $1.97 \pm 0.04$ \\
             &  & TE$_\text{110}$ & $11200 \pm 80$ & $5.80 \pm 0.02$ \\
             &  & TM$_\text{012}$ & $8160 \pm 50$ & $2.83 \pm 0.02$ \\
            \hline
             &  & TM$_\text{010}$ & $163000 \pm 4000$ & $24.7 \pm 0.3$ \\
             &  & TM$_\text{011}$ & $50800 \pm 500$ & $19.54 \pm 0.03$ \\
            0.567> & $\mathrm{Nb}_3\mathrm{Sn}$, $\mathrm{Nb}$, $\mathrm{Al}$ & TE$_\text{111}$ & $154000 \pm 7000$  & $34 \pm 1$ \\
             &  & TE$_\text{110}$ & $68000 \pm 1000$ & $35.2 \pm 0.2$\\
             &  & TM$_\text{012}$ & $85900 \pm 900$ & $29.826 \pm 0.1$   
        \end{tabular}
        \caption{Cavity temperatures $T_{\text{cav}}$ and unloaded quality factors $Q$ of the resonant modes detailed in Table \ref{tab:ModeFreqList} between superconducting transitions during dilution refrigerator cooling, and the factor improvement vs. the no SC material state $Q/Q_i$ with $Q_i$ the initial no SC material $Q$. The corresponding critical temperatures are given in Table \ref{tab:CoolingCritTemps}.}
        \label{tab:CoolingQs}
    \end{table}
    
    \par The two next-lowest frequency resonances see slightly more improvement, at a factor of 2.17, whereas the two highest-frequency modes, $\rm TM_{012}$ and $\rm TE_{111}$, see a small decrease (factor 0.94) and a very slight increase (factor 1.09), respectively. The behavior of these two higher-order modes is also more complex (Figure \ref{fig:CoolingQsByTemp}), showing an initial jump in $Q$ at the $\mathrm{Nb_{3}Sn}$ transition, followed by a drop around $10 \,\rm K$, which then recovers towards the second (Nb) SC transition. As the reported $Q$s are taken as the weighted mean between the two transitions, the lower fractional improvement of these two modes is partially due to this dip lowering the overall average. This behavior is not seen for the three lower-frequency modes and is also not reflected in the coupling factor of these modes, making a mechanical change in the antenna or receiver chain unlikely. The most probable cause is proximity interactions between thinner regions of the $\mathrm{Nb_{3}Sn}$ coating and the $\mathrm{Nb}$ bulk, while the latter is still in its normal state. The drop may then be caused by regions of the $\mathrm{Nb}$ bulk seeing faster cooling due to being more exposed or due to small variations in the exact critical temperature such that some amount of the $\mathrm{Nb}$ bulk begins to superconduct earlier. The interactions between the thin film and varying regions of $\mathrm{Nb}$ in its normal and superconducting state may then become more complex causing regions of the thin film to degrade slightly or otherwise causing regions of higher surface resistance and thereby resulting in the sudden drop in $Q$. As the superconductivity of the $\mathrm{Nb}$ becomes more uniform this then recovers until the full $\mathrm{Nb}$ critical temperature point, where the tuning rod bulk becomes fully superconducting, and the $Q$s see the full transition to the higher $Q$ point. As higher frequency modes with correspondingly more complex field structure, the $\rm TM_{012}$ and $\rm TE_{111}$, may be more affected by these coupling effects than the lower order modes. An alternative explanation is from thermal effects, where some transient heating causes an initial drop in $Q$ that then dissipates at a different rate in regions of the rod near defects, with the higher frequency modes potentially  concentrating more field near these regions and so seeing a more pronounced effect. Note though that no obvious jump or change in rate of temperature was observed to correlate (Figure \ref{fig:CoolingQsByTime}). 
    
    Even considering only data before this drop in $Q$, however, the resulting factor increases are still notably lower than the lower frequency modes with factor improvements of $\approx 1.04$ and $\approx 1.18$ for the $\rm TE_{110}$ and $\rm TM_{012}$, respectively. This aligns well with the geometric factor predictions of Table \ref{tab:sidecarGfactors}, with these two higher order modes having the largest tuning rod geometric factors. Because $1/Q_0 = \sum_i R_{s,i}/G_i$ summed over the cavity surfaces, a mode with a larger sub-surface geometric factor $G_i$ has that sub-surface contribute proportionally less to the total loss. That is, because these higher order modes have comparatively weaker field overlap with the rod surface means that the reduction in rod surface resistance as $\mathrm{Nb_{3}Sn}$ transitions has a smaller effect on the increase in $Q$.
     
    \par The remaining $\mathrm{Nb}$ and $\mathrm{Al}$ transitions see largely similar behavior. For the former, again $Q$s generally improve with the target $\text{TM}_{010}$ mode seeing a factor 2.87 total improvement due to both this and the effect of the $\mathrm{Nb}_3\mathrm{Sn}$ coating. This $Q$ of 18,900 actually exceeds the mean expectation from Table \ref{tab:Q_est}, which makes Equation \ref{eqn:hybridkappaQratio} give a negative resistance for $R_{rod}$ with respect to the aluminum; this is unsurprising given our geometric factor model does not account for the thin or exposed regions of pure niobium that could be present. The $\text{TM}_{011}$ mode sees a decrease in $Q$ with the $\mathrm{Nb}$ transition, such that the overall effect of the superconducting tuning rod has a factor 1.24; this decrease is also seen in the coupling factor, Figure \ref{fig:CoolingCouplings}, implying either the coupling degrades at this transition, resulting in the $Q_0$ decrease, or that this may be due to interactions between the given cavity field and the distribution of SC behavior over the tuning rod, again implying potential defects or regions where the $\mathrm{Nb_3Sn}$ coating is thinner.

    \par Finally, the $\mathrm{Al}$ transition sees large improvements in the $Q$ of all modes. It is unsurprising that the measured $Q$ value is less than the expected value in Table \ref{tab:Q_est}, because the estimate did not capture the residual resistance of the niobium components and is especially sensitive to modeled surface resistance of the copper axle tabs as discussed.

    \subsection{Magnet Ramp Results}
    \label{sec:RampRes}

\begin{figure}
        \centering
        \includegraphics[width=\linewidth]{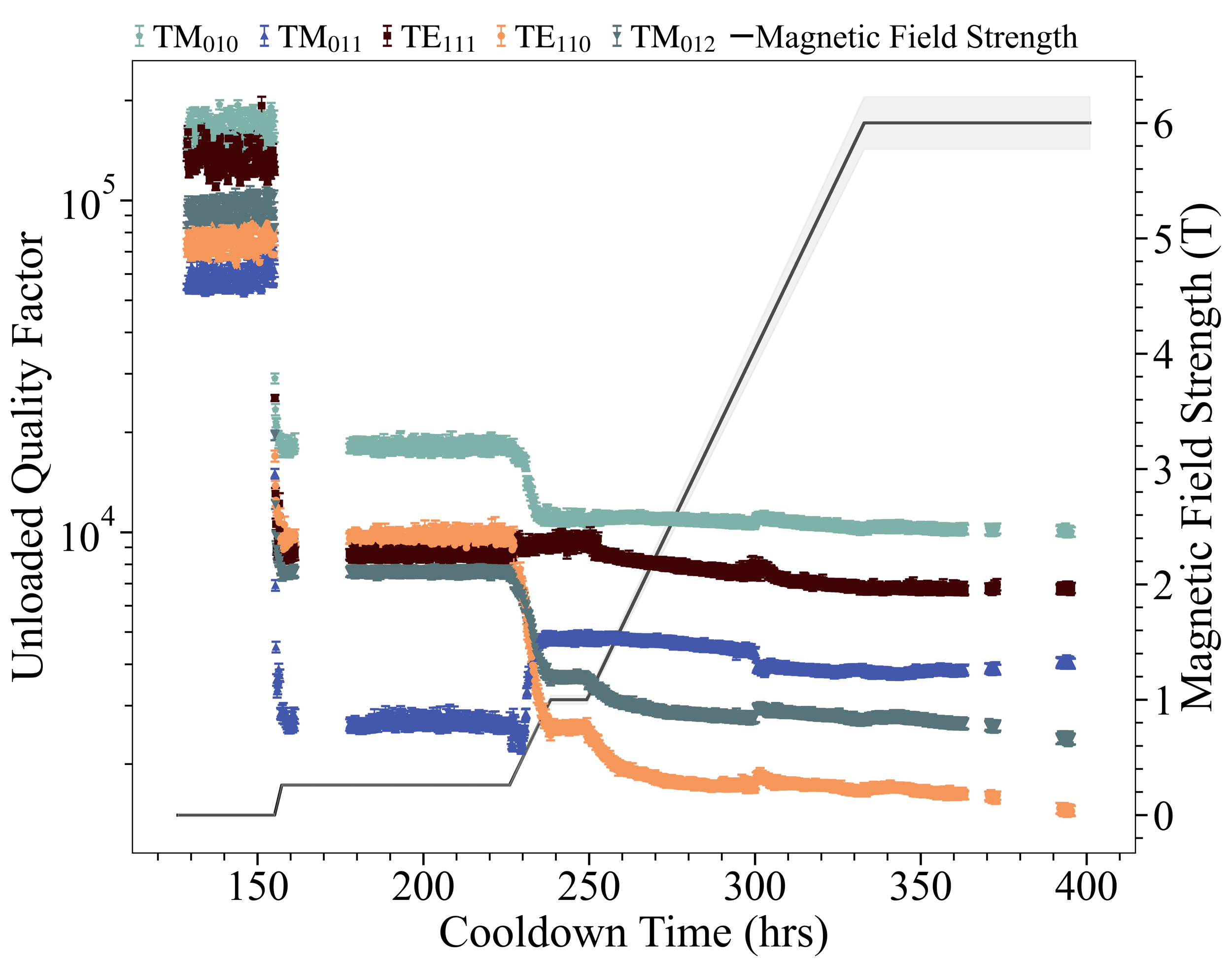}
        \caption{Resonant mode $Q$s (Table \ref{tab:ModeFreqList}) and magnetic field strength (evaluated as the simulated mean field over the cavity volume for the measured nominal field strength) during magnet ramp tests. The gaps in peak data are due to DAQ crashes during the run, ramping was paused during these times and no data of interest is likely to have been lost. The ramping itself was paused twice to maintain temperatures well below any potential critical points and to allow the system to recover from notable temperature jumps (Figure \ref{fig:Rampingtemps}) as the materials quenched. Despite the quenching of the aluminum cavity and niobium tuning rod bulk, and the degradation of the superconducting behavior of the $\mathrm{Nb}_3\mathrm{Sn}$ coating, a notable gain in the cavity $Q$s remain at full field (relative to the completely normal state) as summarized in Table \ref{tab:RampingQs}.}
        \label{fig:RampingFullSyst}
    \end{figure}

    The behavior during magnet ramping tests is significantly more complicated than that during cooling, with individual transitions harder to distinguish. This is unsurprising, as the vortex motion of the niobium components is dependent on the angle of the external field with the normal vector of the surface. Additionally, the unexpected pure niobium transition, whether due to coating imperfections or residual flux expulsion, further complicates disentangling the transitional behavior. The $Q$ evolution and field ramping are shown in Figure \ref{fig:RampingFullSyst}, the gaps in plots in this Section were due to DAQ crashes, all of which occurred whilst the system was in a stationary state (i.e. not whilst ramping) such that no critical data is likely to have been lost. 
    
  \par
\begin{figure}
        \centering
        \includegraphics[width=\linewidth]{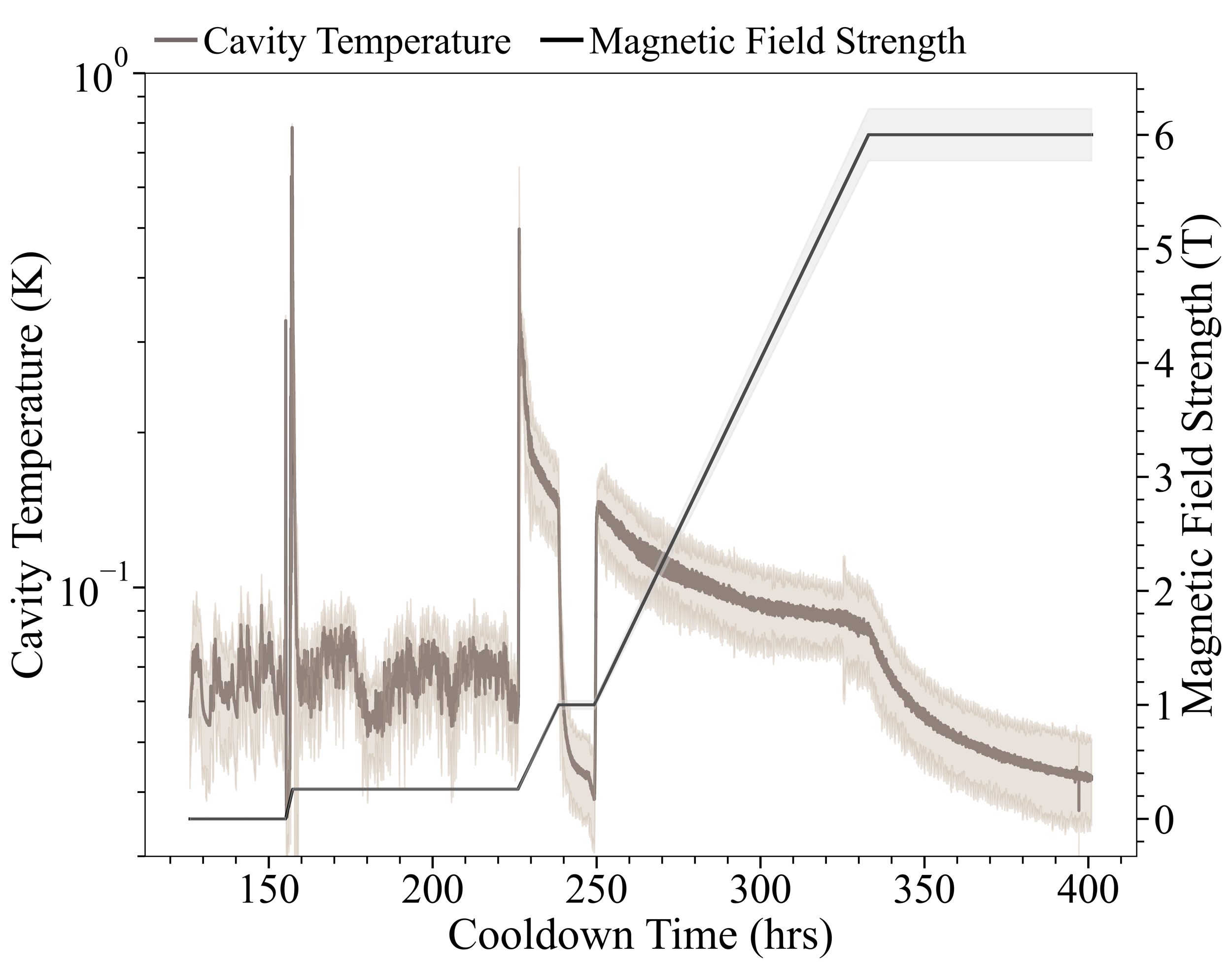}
        \caption{The temperature and field behavior over the ramping tests of Figure \ref{fig:RampingFullSyst}, showing the sharp spikes in temperature as the $\mathrm{Al}$ cavity quenched and during the degradation of superconductivity of $\mathrm{Nb}$ and $\mathrm{Nb_{3}Sn}$.}
        \label{fig:Rampingtemps}
    \end{figure}

 Ramping was paused twice before full field at $6\,\mathrm{T}$, as described in Section \ref{sec:dataAcq}, to allow the system to settle after large temperature jumps at the cavity caused by quenches, which were notably larger than temperature increases generally seen during ramping. The temperature evolution during ramping tests is shown in Figure \ref{fig:Rampingtemps}. This heating is due to both the sudden dissipation of current into the surrounding system and the introduction of a resistive element as the given material becomes either normally conductive or mixed state. The ramping rate was decreased after the first pause to limit heating and temperatures were kept well below the observed critical temperatures of the $\mathrm{Nb}$ and $\mathrm{Nb}_3\mathrm{Sn}$ elements observed during cooling, see Table \ref{tab:CoolingCritTemps}. However, the varying temperature over the run still makes differentiation between field and temperature effects difficult and the observed evolution of the $Q$s immediately after quenching are therefore likely due to both the expected degradation of SC behavior as well as changes in the temperature of the cavity. This is especially true as the poor thermal conductance of the sapphire axles means the tuning rod itself takes a longer time to cool and with no direct measurement of this temperature. 
    
\begin{figure}
        \centering
        \includegraphics[width=\linewidth]{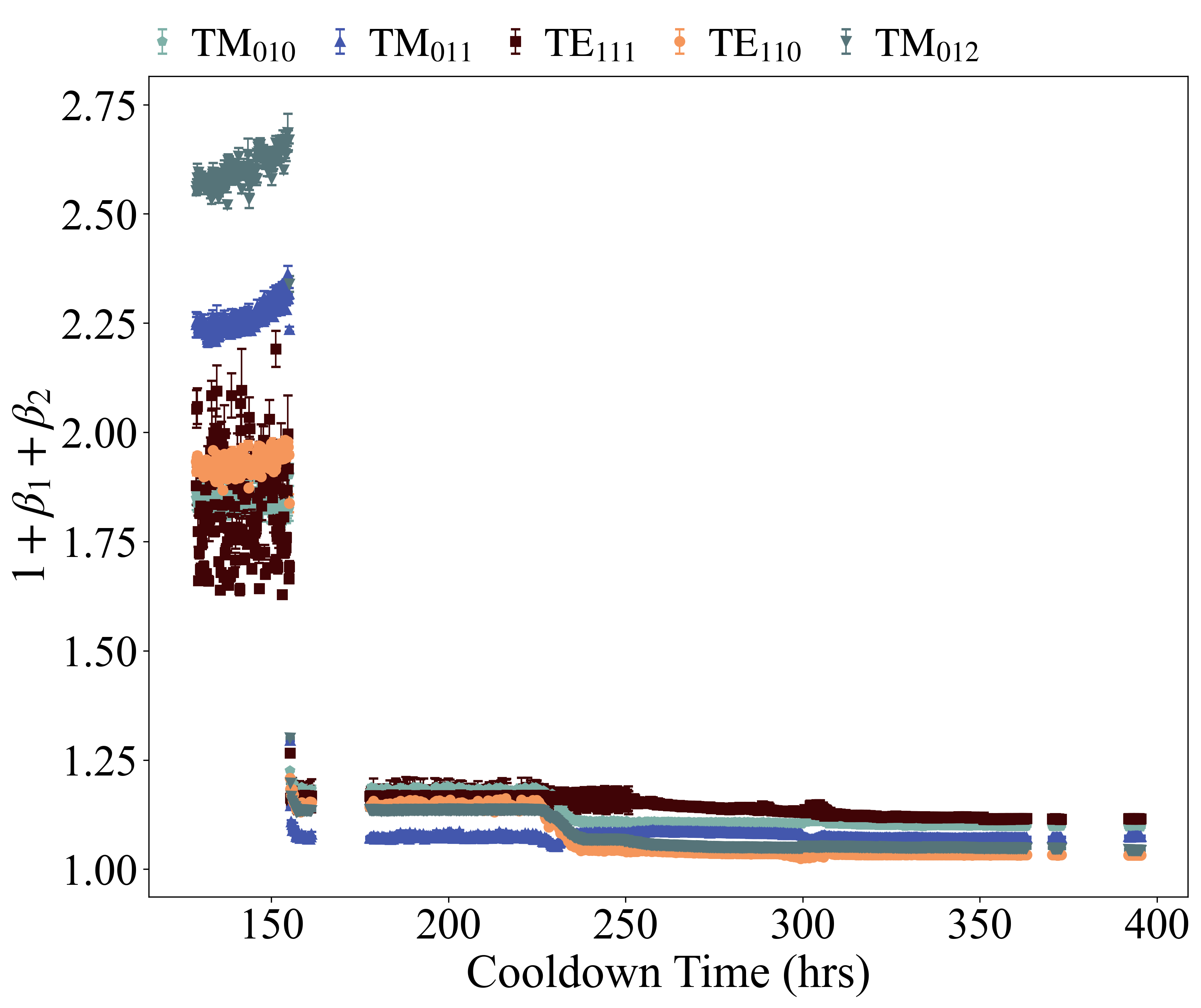}
        \caption{The coupling correction factor for the resonances detailed in Table \ref{tab:ModeFreqList} during magnet ramping tests, calculation details are as in Section \ref{sec:dataAcq}.}
        \label{fig:RampingCoupling}
    \end{figure}

    The coupling correction factor during ramping tests is shown in Figure \ref{fig:RampingCoupling} and largely follows the same trend as in Section \ref{sec:CoolRes} with $\beta$ values tracking the evolution of $Q$ as expected and variation outside of that attributed to SC interactions at rod defects and to temperature variation over the run.

    \par The initial quenching of the $\mathrm{Al}$ cavity occurred almost immediately as ramping was started, and resulted in a sharp drop in all measured $Q$s. The slower decline in $Q$s occurring after ramping was resumed is likely due to the $\mathrm{Nb}_3\mathrm{Sn}$ tuning rod coating entering a mixed state, beyond its $H_{c1}$ critical field, resulting in a slower degradation of superconducting behavior. This implies that, despite any exposed regions of $\mathrm{Nb}$, the $\mathrm{Nb}_3\mathrm{Sn}$ coating still has a significant effect on the cavity performance; this could be quantified more rigorously with simulation analysis similar Reference \onlinecite{PhysRevApplied.20.034004} in a future publication.
    \par The decline in $Q$ then begins to decrease further now due to the transition of the $\mathrm{Nb}$ tuning rod bulk from its Meissner to mixed to normally conductive states, increasing the slope of Q degradation beyond just the $\mathrm{Nb}_3\mathrm{Sn}$ contribution. This could be due to either exposed niobium regions, or the substrate providing less field expulsion, allowing further flux penetration into the coating. Beyond $\approx1.2\,\mathrm{T}$, the $Q$s largely plateau, with the degradation in $Q$ much slower up to full field at $6\,\mathrm{T}$. The jump in $Q$s around 300 hours ($\approx4\,\mathrm{T}$) are expected to be due to the antenna physically jumping, as it is also visible in the coupling factor plots.
    
    \par
\begin{figure}[ht]
        \centering
        \includegraphics[width=\linewidth]{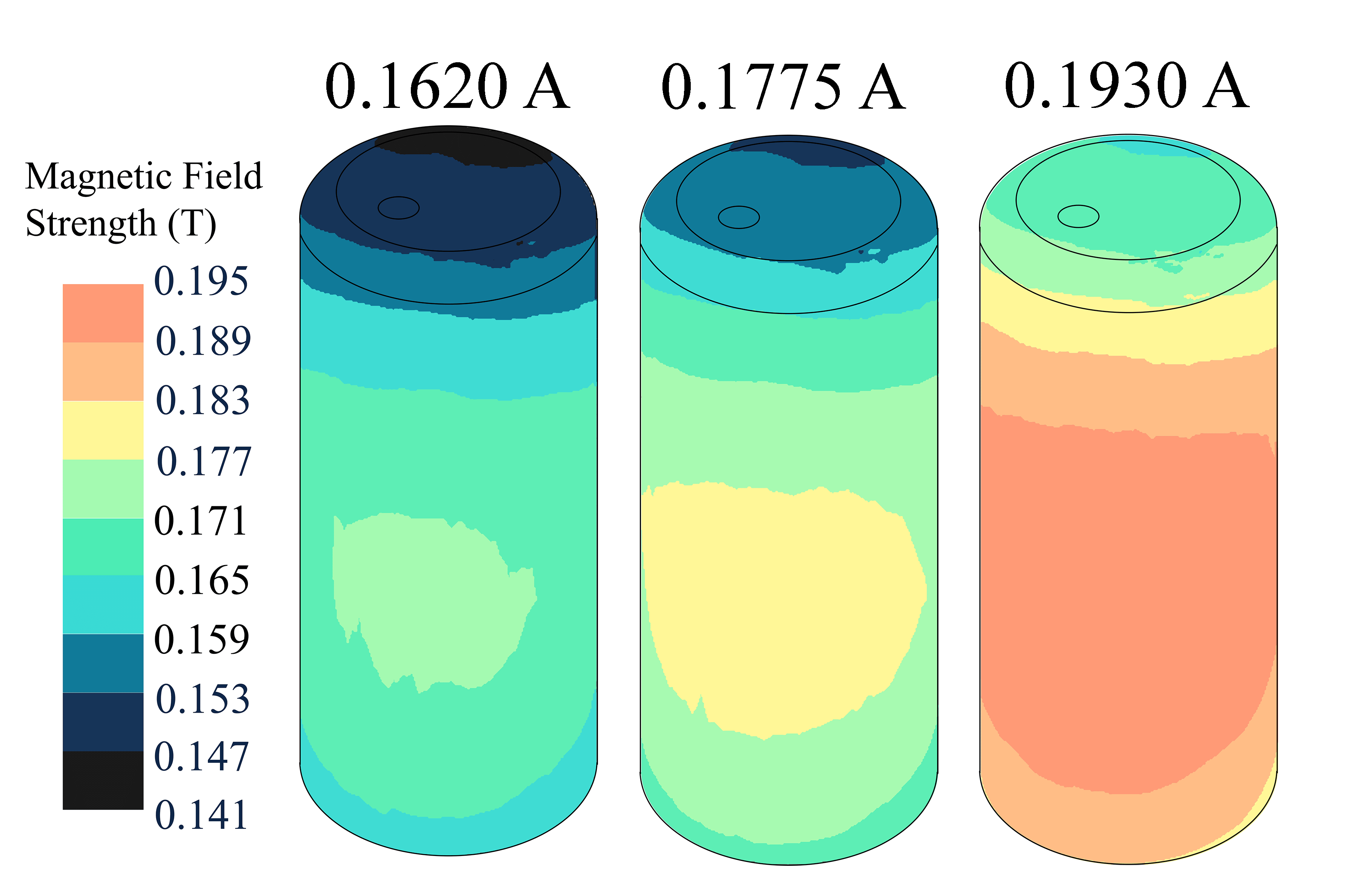}
        \caption{ANSYS Maxwell 3D simulations of the evolution of the magnetic field magnitude at the surface of the tuning rod during ramping of the field around the $H_{c1}=0.17\,\text{T}$ critical field of niobium. These are for the tuning rod held at 140 degrees (with 0 degrees the position closest to the wall and 180 closets to the center).}
        \label{fig:TuningRodFieldEvolu}
    \end{figure}

 Exact critical fields are hard to extract due to both the $\mathrm{Nb}$ and $\mathrm{Nb}_3\mathrm{Sn}$ components existing in a mixed state at the same time, as well as the aforementioned temperature and coupling effects. Additionally, as the tuning rod and cavity are mounted off-center in the magnet's internal bore, the field experienced is not uniform. Table \ref{tab:CritFieldvsNomB} shows the expected critical fields for each material and the nominal field strengths at which any part and at which every part of the tuning rod experiences the expected critical field, referred to as the minimum and maximum nominal field, respectively. Figure \ref{fig:TuningRodFieldEvolu} shows the evolution of the field experienced at the surface of the tuning rod around the $H_{c1}$ critical field of $\mathrm{Nb}$ at $0.17\,\mathrm{T}$ to illustrate this variation in experienced field. The observed critical fields were extracted from Figure \ref{fig:RampingFullSyst} using regimes of consistent $Q$ evolution as well as the temperature and simulation data to identify the critical fields, these are also given in Table \ref{tab:CritFieldvsNomB}. The observed critical fields are generally slightly higher than these simulated ranges, but are of similar order; this is expected given the non-uniform field over the rod, the gradual nature of the mixed-state transitions, and the temperature dependence of the critical fields, such that the field at which a $Q$ change becomes visible often does not coincide with the nominal literature value.
    
    \begin{table}[ht]
        \centering
        \begin{tabular}{c|c|c|c|c}
            Transition & Exp $H_{c}$ & $\text{min}(B_{\text{nom}})$ & $\text{max}(B_{\text{nom}})$ & Obs $H_{c}$ \\
            \hline
            \hline
            $\mathrm{Al}$ $H_{c}$ & 0.01 & 0.0095557 & 0.01134 & 0 \\
            $\mathrm{Nb}_3\mathrm{Sn}$ $H_{c1}$ & 0.038 & 0.036312 & 0.04311  &  0.00541 -- 0.00561 \\
            $\mathrm{Nb}$ $H_{c1}$ & 0.17 & 0.16245 & 0.19285 & 0.233 -- 0.318 \\
            $\mathrm{Nb}$ $H_{c2}$ & 0.4 & 0.38223 & 0.45377 & 0.557 -- 0.976
        \end{tabular}
        \caption{The expected critical fields for the materials used in the cavity (see Figure \ref{fig:CavitySchematic}) and the nominal magnet field strengths at which any $\text{min}(B_{\text{nom}})$ and all $\text{max}(B_{\text{nom}})$ of the relevant component experiences that critical field, considering the cavity inner wall for $\mathrm{Al}$ and the tuning rod outer wall otherwise. Obs $H_{c}$ give ranges of field strengths for which changes in $Q$ were seen that are likely due to the corresponding transition. \\
        The thickness of the $\mathrm{Nb}_3\mathrm{Sn}$ coating is taken as a negligible effect compared to other simulation and positioning uncertainties (e.g. variations in current in the coils and resulting field shape, and the exact vertical placement and rotation of the tuning rod), so that values for the $\mathrm{Nb}$ bulk are evaluated on the surface of the tuning rod identically to the $\mathrm{Nb}_3\mathrm{Sn}$ case. Simulations are done using coil current information provided by Oxford Instruments and in ANSYS HFSS Maxwell 3D \cite{ansyselectromagnetics}. A schematic of the variation in field experience by the tuning rod over an example critical field is shown in Figure \ref{fig:TuningRodFieldEvolu}. $\mathrm{Nb_3Sn}\ H_{c2}$ is omitted as it is far beyond the magnet max field at $30\,\rm T$. Critical field values are taken from Reference \onlinecite{CritFieldofSCAl} for $\mathrm{Al}$ and Reference \onlinecite{Nb3SnForCavs} otherwise.}
        \label{tab:CritFieldvsNomB}
    \end{table}

    Table \ref{tab:RampingQs} shows the $Q$s measured during cooling with no superconducting materials and with only $\mathrm{Nb}_3\mathrm{Sn}$ superconducting, both taken from the cooling results, as well as the measured $Q$s at full field.

    \begin{table}
        \centering
        \begin{tabular}{c|c|c|c}
             SC Materials & Peak & $Q$ & $Q/Q_{i}$  \\
            \hline
            \hline
             & TM$_\text{010}$ & $6590 \pm 80$ & 1 \\
             & TM$_\text{011}$ &  $2600 \pm 30$ & 1\\
             None & TE$_\text{111}$ & $4570 \pm 60$ & 1 \\
             & TE$_\text{110}$ & $1930 \pm 20$ & 1 \\
             & TM$_\text{012}$ & $2880 \pm 40$ & 1 \\             
            \hline
              & TM$_\text{010}$ &  $12260 \pm 60$ & $1.86\pm 0.02$\\
              & TM$_\text{011}$ &  $5630 \pm 30$ & $2.17 \pm 0.03$ \\
            $\mathrm{Nb}_3\mathrm{Sn}$ & TE$_\text{111}$  & $9900 \pm 50$ & $2.17 \pm 0.03$ \\
             & TE$_\text{110}$ & $1820 \pm 40$ & $0.94 \pm 0.02$ \\
              & TM$_\text{012}$ & $3140 \pm 20$ & $1.09 \pm 0.02$ \\
            \hline
             & TM$_\text{010}$  & $10260 \pm 20$ & $1.56 \pm 0.02$ \\
              & TM$_\text{011}$  & $3824 \pm 9$ & $1.47 \pm 0.02$ \\
            $\mathrm{Nb}_3\mathrm{Sn}, \, 6\,\text{T}$ & TE$_\text{111}$ & $6770 \pm 20$ & $1.48 \pm 0.02$ \\
              & TE$_\text{110}$ &  $1639 \pm 7$ & $0.849 \pm 0.009$ \\
              & TM$_\text{012}$ &  $2675 \pm 9$ & $0.93 \pm 0.01$ 
        \end{tabular}
        \caption{Unloaded quality factors $Q_{0}$ of the resonances detailed in Table \ref{tab:ModeFreqList} with the dilution refrigerator during cooling before any materials transitioned to their superconducting states, during cooling when only the  $\mathrm{Nb}_3\mathrm{Sn}$ tuning rod coating was superconducting, and at full field with degraded $\mathrm{Nb}_3\mathrm{Sn}$ superconductivity. $Q/Q_{i}$ is the factor change in unloaded $Q$ from the fully normal state.}
        \label{tab:RampingQs}
    \end{table}
    
    Even at full field, a notable improvement in the $Q$s of most modes remains. Though the higher frequency $\text{TE}_{110}$ and $\text{TM}_{012}$ modes see a slight degradation in $Q$, these degradations (as discussed in Section \ref{sec:CoolRes}) are likely due to interactions between the mixed state $\mathrm{Nb_{3}Sn}$ and normal $\mathrm{Nb}$ bulk.
    \par Most notably, the target mode for most axion experiments, the $\text{TM}_{010}$ mode sees a 1.56 times improvement at full field from the fully normal phase. This is also a 1.27 times improvement of the expected $Q$ value of an entirely aluminum cavity system as shown in Table \ref{tab:Q_est}. By virtue of Equation \ref{eqn:hybridkappaQratio}, this implies that $R_{rod}=7.88\,\rm m\Omega=0.47\cdot R_{\mathrm{6061Al}}$, but unfortunately would be $2.33\cdot R_{\mathrm{Cu}}$.
    
\section{Conclusion}
\par An aluminum clamshell microwave resonator loaded with a niobium tuning rod coated with $\mathrm{Nb_3Sn}$, known as the ADMX Sidecar, was characterized in a dilution fridge through its critical temperature transitions down to $25\,\mathrm{mK}$ and through magnetic fields up to $6\,\mathrm{T}$. The tuning rod was held at a fixed position throughout the measurements, corresponding to a resonant frequency of $5.2\,\mathrm{GHz}$ for the TM$_\text{010}$ mode, the standard mode that is monitored during axion searches. The quality factor of five resonant modes of the cavity system were monitored throughout the process, including the TM$_\text{010}$.
\par Distinct superconducting transitions were observed near the expected critical temperatures of $\mathrm{Nb_3Sn}$, pure niobium, and aluminum. The observed transition of pure niobium might imply that the rod may have imperfections in the coating. Another explanation is the expulsion of residual flux in the coating by the substrate's Meissner transition. Nonetheless, the TM$_\text{010}$ mode exhibited an improvement factor of 1.86 and 2.87 through the two niobium transitions, and implied the rod had a surface resistance of $\approx16\%$ that of the normal aluminum and $79\%$ that of copper at cryogenic temperatures. The TM$_\text{010}$ mode after the aluminum transition had a $Q$ of 163,000. 
\par Once the magnet was turned on, the aluminum immediately transitioned back to its normal state producing a sharp decline in the quality factor of all modes. The quality factors observably degraded gradually with field after this transition, which is indicative of the $\mathrm{Nb_3Sn}$ coating being in its mixed state with increasing fluxon density as expected. The quality factors mostly plateaued above $1.2\,\mathrm{T}$, nonetheless the TM$_\text{010}$ mode still showed an improvement factor of 1.56 over its normal state $Q$ at the full field of $6\,\mathrm{T}$ and implied the tuning rod had a resistance $47\%$ that of the aluminum. However, this translates to a resistance 2.33 times that of copper, which does not bode well for axion searches that operate in excess of $6\,\mathrm{T}$. 
\par This higher observed resistance could be due to the presence of exposed pure niobium which would have a $H_{c2}\approx0.4\,\rm T$ and de-pinning frequency of $\nu _p \approx 673\,\rm MHz$ \cite{NbPinningFreq,PhysRev.139.A1515,PhysRev.139.A1515,PhysRevAccelBeams.23.023102}, making these regions normal with a resistance approximately 24 times that of cryogenic copper. Furthermore, even if these exposed regions are somehow still in a mixed state, $B_{ext}<0.4\,\rm T$, vortex motion would be highly dissipative at $5.2\,\rm GHz$, far above pure niobium's de-pinning frequency. In contrast, $\rm Nb_3Sn$ should be operating in a mixed state below its de-pinning frequency with less dissipation. Surface characterization of the rod via microscopy or spectrographic methods may corroborate this theory. 
\par Alternatively, the breakdown of the Meissner effect in the niobium substrate as the field is increased would expose the coating to more flux penetration, increasing its resistance at a higher rate. Simulations of the angular dependency of superconducting surface currents within the non-uniform field may also better explain the degradation in $Q$ seen during magnet measurements\cite{PhysRevApplied.20.034004}.
\par These measurements inform ongoing simulation work of hybrid superconducting cavity resonator systems, which constitutes a publication of its own. Additionally, a $\mathrm{Cu-Nb_3Sn}$ ADMX Sidecar cavity was used for an axion-like particle search within the ADMX insert; the science results of this search will also be reported in a separate publication.
\par This work informs the future development of magnetic field tolerant superconducting cavities for high frequency axion searches. Based on our findings, $\mathrm{Nb_3Sn}$ may still improve the quality factor over standard copper cavity haloscopes, but using a pure niobium substrate must be done with careful handling; coating imperfections may quickly spoil any improvement due to niobium's higher resistance. Alternative coating processes, such as ones that use a bronze substrate\cite{Withanage_2021}, or alternative superconductors, such as REBCO materials, specifically YBCO tapes\cite{Ahyoune2025,742y-g3fd,CAPPPatras,CAPPYBCO1,CAPPYBCO2,CAPPYBCO3}, may also show promise for this application.

% If you have acknowledgments, this puts in the proper section head.
\begin{acknowledgments}
This work is supported by U.S. Department of Energy through Awards No DE-SC0009800, No. DE-SC0009723, No. DE-SC0010296, No. DE-SC0010280, No. DE-SC0011665, No. DEFG02-97ER41029, No. DE-FG02-96ER40956, No. DE-C03-76SF00098, No. DE-SC-0022148 and No. DE-SC0017987. Additional support was provided by the Heising-Simons Foundation and by the Lawrence Livermore National Laboratory LDRD office. PNNL is a multi-program national laboratory operated for the U.S. DOE by Battelle Memorial Institute under Contract No. DE-AC05-76RL01830. The Sheffield group acknowledges support from the UK Science and Technology Facilities Council (STFC) under grants ST/ T006811/1 and ST/X005879/1. This work was performed under the auspices of the U.S. DOE by Lawrence Livermore National Laboratory under Contract DE-AC52-07NA27344. The work was supported by the U.S. Department of Energy, Office of Science, National Quantum Information Science Research Centers, Superconducting Quantum Materials and Systems (SQMS) Center under the contract No. DE-AC02-07CH11359. This research made use of scikit-rf, an open-source Python package for RF and Microwave applications, as well as COMSOL Multiphysics® v. 6.4. LLNL Release No. LLNL-JRNL-2024695.
\end{acknowledgments}

\section*{Data Availability Statement}
Data are available from the authors upon reasonable request and with the permission of Pacific Northwest National Laboratory, Lawrence Livermore National Laboratory, and the University of Sheffield.

% Create the reference section using BibTeX:
\bibliography{bibliography.bib}

\appendix

\section{Spectra Fitting and Coupling Analysis}
\label{app:SpectandCoup}

\subsection{Spectra Fitting}
\begin{figure*}
        \centering
        \includegraphics[width=\linewidth]{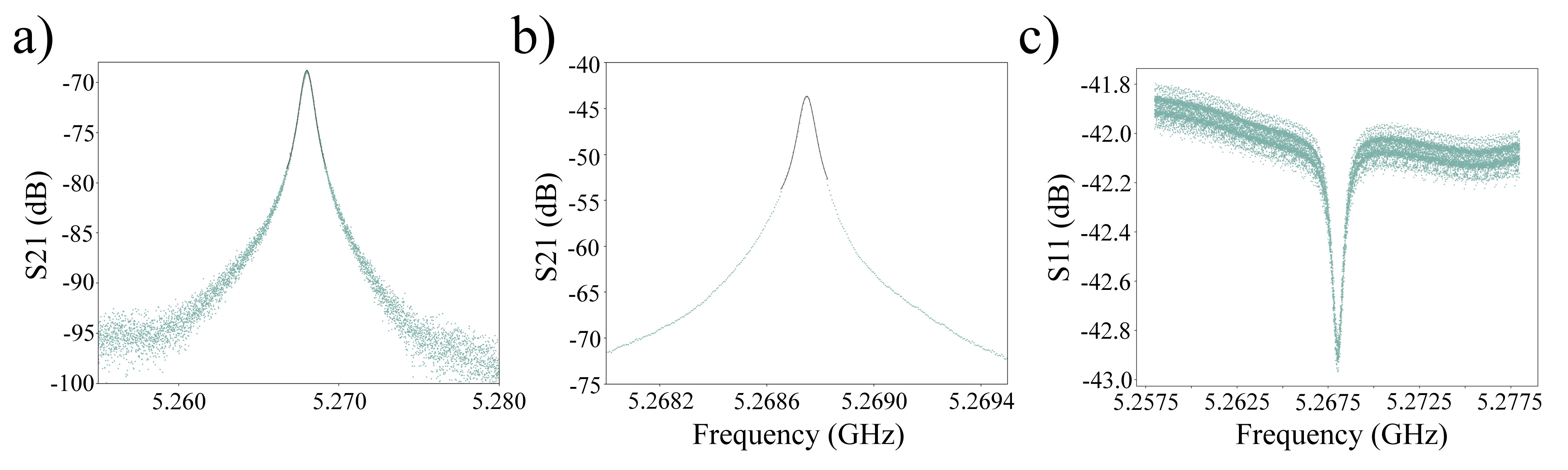}
        \caption{(a) and (b): Transmission ($S_{21}$) spectra for the $\text{TM}_{010}$ peak, showing a narrow frequency region around the resonant frequency, with brown solid line the Lorentzian fit as in Equation \ref{eq:LorCav}. These are for (a). spectra taken around $40\,\rm K$ during cooling, with no superconducting materials and the lowest $Q$, and (b). spectra taken around 8.5 mK immediately before magnet ramping with all materials superconducting in the Meissner state and the highest $Q$. (c): An example reflection ($S_{11}$) measurement showing the observed vertical striping described in the main text.}
        \label{fig:TM010Spect}
    \end{figure*}
Measurements of unloaded quality factor here were made using the equations given in Section \ref{sec:dataAcq}. The $S_{21}$, transmission, data for all peaks was analyzed by fitting to a Lorentzian as in Equation \ref{eq:LorCav}. Example transmission resonant peaks are shown for the target $\text{TM}_{010}$ mode in Figure \ref{fig:TM010Spect}(a) and (b), with no superconducting materials and at base dilution refrigerator temperature and zero field with all materials superconducting; the increase in $Q$ and corresponding narrowing of the peak bandwidth are clearly visible. 
    
\par When fitting transmission resonances, especially those with high $Q$ and narrow bandwidths which have fewer data points about the resonant frequency, the amount of data about the resonance considered can have a notable effect on the resulting fit parameters. Errors were so quantified by fitting data within an $8\, \rm dB$ drop from the amplitude at the resonant frequency with this result given as the quoted values throughout, fits were then repeated with a wider $14\, \rm dB$ drop and the uncertainty in the loaded $Q_L=\nu_0/\Gamma$ based on the difference in the resulting fit parameters.
    
\par The calculation of the unloaded quality factor then required calculations of $\beta_{1}$ and $\beta_{2}$ from the reflection coefficients. Figure \ref{fig:TM010Spect}(c) shows an example $S_{11}$ spectra for the $\text{TM}_{010}$ modes. The reflection measurements $\rm S_{11}$ and $\rm S_{22}$ were subject to instrumental artifacts that manifested in the shown vertical striping in the results. These artifacts were minimized in the analysis by filtering to retain only the denser-sampled data near the spectrum center. An inverse Lorentzian was then fitted to the remainder to determine the amplitude. The noise floor was found by removing the peak, fitting the remaining data to an order 2 polynomial, and taking the polynomial value at the resonant frequency; for very low couplings these fits were checked manually. Uncertainties were taken as the standard deviation of the unfiltered data in a small region around the peak, combined with the standard deviation of the filtered data about the fitted polynomial.

\begin{figure*}
        \centering
        \includegraphics[width=\linewidth]{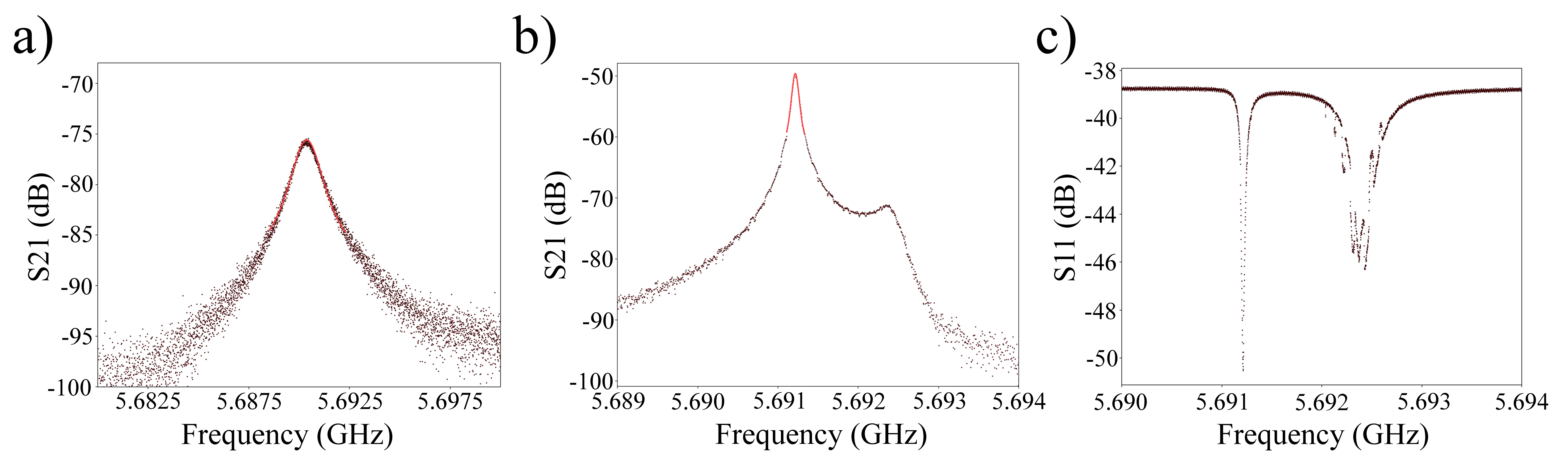}
        \caption{(a) and (b): Transmission (S21) spectra for the $\text{TE}_{111}$ degenerate mode pair, showing a narrow frequency region around the resonant frequency and with solid red line a Lorentzian fit as in Equation \ref{eq:LorCav}. These for (a): a spectra taken around $40\,\rm K$ during cooling, with no superconducting materials and with the degenerate pair forming a single broader feature though skewed to lower frequency, and (b): a spectra taken around $8.5\, \rm mK$ immediately before magnet ramping with all materials superconducting with the two peaks clearly separable, with (c) the corresponding reflection (S11) spectrum for (b). }
        \label{fig:TEPairTrans}
    \end{figure*}

Of note is that the $\rm TE_{111}$ mode saw notable splitting when $Q$s were high. Figure \ref{fig:TEPairTrans} shows example $S_{21}$ transmission plots at low and high $Q$ and an example reflection plot showing this degenerate splitting behavior. Where the spectra formed two distinct peaks the transmission peak fits were performed over the lower frequency peak (as that with the higher amplitude) and the reflection again over the lower which saw correspondingly deeper troughs. All TE modes constitute a degenerate mode pair such that it is unclear whether the increase in SC behavior resulted in further degeneracy lifting such that the two peaks became separable or whether the inherent splitting of the two is enough that the peaks simply become visually separable when narrow enough. From comparison to the other TE ($\rm TE_{110}$) mode where this splitting was not seen, the $Q$ is indeed higher for the split $\rm TE_{111}$ mode, such that this does not necessarily imply this peak sees more degeneracy lifting than the other TE mode. The relevance of this to the analysis here is that this represents a higher uncertainty in the reported $Q$s of this TE mode, which have been factored in to the given errors as much as possible. Further uncertainty which is harder to quantify from fits alone may be caused where, at lower $Q$s, the two modes cannot be distinguished resulting in artificially higher bandwidth and so an artificially lower $Q$. Also of note are cases where coupling differences between the two ports meant that distinct peaks were visible only for one reflection S-parameter, though here as the depth of the trough is the relevant parameter and the measurements were always taken over the deeper dip, this should not result in any notable change and is mentioned here only for completeness.

This peak does see notably higher spread in measured $Q$ especially at high $Q$ (see Sections \ref{sec:CoolRes} and \ref{sec:RampRes}), this is likely due predominantly to the mode crossing-like behavior between the two degenerate modes with their fields and interactions with the SC components of the cavity being somewhat coupled and therefore affected by the other. This mode was also one of those that saw a decrease in $Q$ with the $\mathrm{Nb}$ transition, which may also be in part due to the hybridized and more complex field structures of the degenerate pair resulting in potential defects in the rod coating having a larger effect.

\subsubsection{Coupling Analysis}
\label{app:couplingphase}

\begin{figure*}
        \centering
        \includegraphics[width=\linewidth]{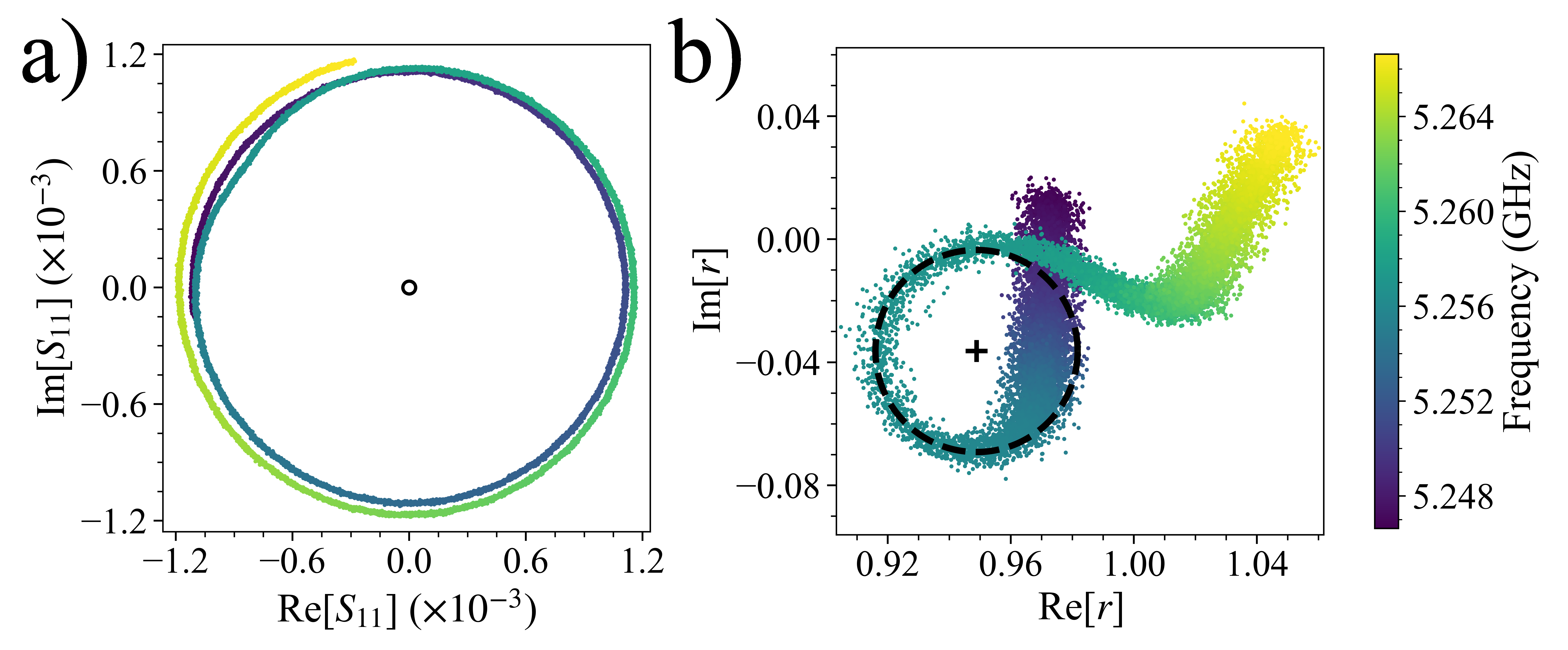}
        \caption{An example fit to determine coupling regime from complex reflection S-parameters, here for $S_{11}$ data from the $\rm TM_{010}$ mode taken with no materials superconducting. (a): raw complex $S_{11}$ data and (b) data after phase delay removal and normalization such that the remaining arc is predominantly due to the resonance and the off-resonance point is ideally $r = 1 + 0i$, the black line shows the performed circle fit with undercoupling condition that the origin is not enclosed or crossed by this circle.}
        \label{fig:coupfit}
    \end{figure*}

To verify that the measured resonances were all in the undercoupled regime on both ports, phase fits were performed of the two reflection scattering parameters ($S_{11}$ and $S_{22}$). The relevance of this here is that equations \ref{eq:beta1} and \ref{eq:beta2}, used to convert from the loaded to unloaded quality factor, slightly vary in sign in the case where the ports are either side of critically coupled (i.e. where one is overcoupled and the other under). An example fit is shown in Figure \ref{fig:coupfit}.

The raw complex trace $S_{11}(\nu)$ is distorted by a frequency-dependent cable phase delay $\tau$, which winds the trace into a spiral in the complex  plane. To remove this, the corrected trace is formed as

\begin{equation}
    \tilde{S}_{11}(\nu) = S_{11}(\nu)\,e^{2\pi i \nu \tau},
\end{equation}

where $\tau$ is estimated from the slope of the unwrapped phase in the off-resonance tails of the spectrum, far from the resonant frequency. Linear fits to the unwrapped phase $\phi(\nu) = \arg[S_{11}(\nu)]$ in the upper and lower frequency tails give slopes $\mathrm{d}\phi/\mathrm{d}\nu$, from which $\tau = -\langle\mathrm{d}\phi/\mathrm{d}\nu\rangle / 2\pi$.

After delay removal, $\tilde{S}_{11}(\nu)$ ideally traces a circle in the complex plane. The coupling regime is itself determined by the location of the origin relative to the circle, with the undercoupled case seen here where the origin lies outside the circle, the overcoupled where it is within it, and critical coupling where the circle crosses through the origin at the resonant frequency. 

Practically, the resulting shapes are somewhat distorted or have tails, as seen in Figure \ref{fig:coupfit}, this is expected to be the result primarily of standing waves in the cable run resulting in variations in the off-resonance phase or remaining unfitted phase delay. The coupling regime was not found to change within the span of these tails. 

A normalization was performed prior to fitting, defining the normalized reflection coefficient as

\begin{equation}
    r(\nu) = \frac{\tilde{S}_{11}(\nu)}{S_{11}^{\mathrm{bg}}},
\end{equation}

where $S_{11}^{\mathrm{bg}}$ is the median of the delay-corrected trace $\tilde{S}_{11}(\nu)$ evaluated in the off-resonance tails, mapping the off-resonance point to $r = 1 + 0i$. This normalization was done to trial extracting the coupling coefficients $\beta_1$ and $\beta_2$ directly from the geometry of the fit circle; however, the noise in the data and the uncertainty in the phase delay removal introduced scatter in the final fit such that quantitative coupling extraction was less reliable than the $S$-parameter reflection method used in the main text. The normalized circle fit was therefore used only as a coupling regime test.

The circle fits were performed by minimizing the radial scatter of the data about a candidate circle, with points weighted by a Lorentzian in the reduced frequency $\epsilon = 2(\nu - \nu_0)/\Gamma$ to down-weight the off-resonance background. 

To verify the fits worked consistently, the fit cable delay was checked to be consistent and in reasonable range. Here, these were found to be $115 \pm 20\,\rm ns$, or $12 \pm 2\,\rm m$ of cabling, which aligns well with the estimated actual one-way total cable length of $11.5\,\rm m$, with the uncertainty on the fit values attributed to standing waves and reflections within the cable run. Some fits failed to produce a clear arc due to noisy data and resultingly unconstrained fits, however, the fits overall returned as undercoupled for the majority of fits in each combination of SC materials, as the antennas were stationary it is therefore assumed that this reflects consistent undercoupling for the full run.

This found that all resonances remained undercoupled throughout the data-taking run, the observed decreases in $Q$ and disagreement with simulation noted in this paper are then expected to be the result of the interaction between the given modes and the SC tuning rod walls and not a result of assumptions made in the analysis.

\end{document}